\pdfoutput=1
\documentclass[11pt]{article}

\usepackage[T1]{fontenc}
\usepackage[utf8]{inputenc}
\usepackage[margin=1in]{geometry}
\usepackage{amsmath,amssymb}
\usepackage{graphicx}
\usepackage{booktabs}
\usepackage{longtable}
\usepackage{array}
\usepackage{calc}
\usepackage{caption}
\usepackage[hidelinks]{hyperref}
\usepackage[section]{placeins} 
\usepackage{flafter} 
\makeatletter
\AtBeginDocument{\expandafter\renewcommand\expandafter\subsection\expandafter
  {\expandafter\@fb@secFB\subsection}}
\makeatother

\providecommand{\tightlist}{%
  \setlength{\itemsep}{0pt}\setlength{\parskip}{0pt}}

\title{Computing the entropy rate of a\\ quantized stationary Gaussian process}
\author{Jeremy Magland\thanks{Center for Computational Mathematics, Flatiron Institute, 162 Fifth Avenue, New York, NY 10010, USA. Email: \href{mailto:jmagland@flatironinstitute.org}{jmagland@flatironinstitute.org}}}
\date{}

\begin{document}

\maketitle

\begin{abstract}
The entropy rate of a uniformly quantized stationary Gaussian process is the limit on how far the quantized signal can be compressed without loss. It has no closed form, and the classical high-resolution formula fails when the spectral density is small on part of the band, as it is after routine smoothing or filtering. Here we present two ways to compute it. The first is an analytical approximation in which the quantization noise acts as a floor on the spectral density. The second is a particle-filter estimate of the exact rate. For quantization steps up to the signal standard deviation, the two agree to within a few millibits per sample in most cases, including processes on which the classical formula fails. At a step of a quarter of the standard deviation, linear predictive coding comes within 2 to 4 percent of the rate, while general-purpose compressors remain 12 to 89 percent above it.

\end{abstract}

\section{Introduction}\label{introduction}

A basic question about any digitized noisy signal is how far it can be compressed without loss. When a continuous-valued signal is digitized by uniform quantization, the resulting integer sequence is what is stored, and its entropy rate is the fundamental limit on lossless compression \cite[Chs.~2 and~4--7]{MacKay2003}: the number of bits per sample the signal actually carries. We take the underlying signal to be a stationary Gaussian process, so the entropy rate is determined by the spectral density and the quantization step. Modeling the signal as pure noise is less restrictive than it may appear: in a noisy measurement the incompressible content is largely the noise rather than the structured underlying signal, so the noise entropy rate accounts for most of the compression cost. The application that motivates this work is extracellular electrophysiology, where multichannel recordings sampled at tens of kilohertz are dominated by filtered noise and the storage cost is a practical concern. The background of such recordings, the summed activity of many distant neurons together with thermal and amplifier noise, is commonly treated as Gaussian \cite{QuianQuiroga2004}; spikes are non-Gaussian but occupy a small fraction of the samples, and stationarity holds approximately over windows short enough for the spectrum to be roughly constant. Moreover, the Gaussian has the largest entropy rate for a given spectrum \cite{CoverThomas2006}, so at fine quantization the computed rate is approximately an upper bound for a non-Gaussian signal with the same spectrum.

Let \(\{X_t\}_{t\in\mathbb Z}\) be a zero-mean stationary Gaussian process with autocovariance \(\gamma_k = \mathbb E[X_t X_{t+k}]\), spectral density \(S(\omega) = \sum_k \gamma_k e^{-ik\omega} \ge 0\) on \([-\pi,\pi]\), and variance \(\sigma^2 = \gamma_0\). The process is quantized with step \(\Delta > 0\):
\[Y_t = \operatorname{round}(X_t/\Delta) \in \mathbb Z,\]
so that \(Y_t = y\) if and only if \(X_t\) lies in the cell \(\mathcal C_y = [(y-\tfrac12)\Delta,\,(y+\tfrac12)\Delta)\). Writing \(Y_1^n = (Y_1,\dots,Y_n)\), and likewise for other sequences, the quantity of interest is the entropy rate, in bits per sample,
\[\bar H = \lim_{n\to\infty}\frac1n H(Y_1^n), \qquad H(Y_1^n) = -\sum_{y\in\mathbb Z^n} P(Y_1^n = y)\,\log_2 P(Y_1^n = y).\]
Although the alphabet is countably infinite, \(\bar H\) is finite because \(X_t\) has finite variance.

The entropy rate is a per-sample quantity, but it is not the entropy \(H(Y_1)\) of a single quantized sample. The latter is the cost of coding each sample on its own, in ignorance of its neighbors. Because \(X_t\) is correlated in time, successive quantized values are partly predictable from their past, and a coder that models this pays less per sample. The rate \(\bar H\) is what remains once all such temporal structure is accounted for, and the gap \(H(Y_1) - \bar H\) measures the compressibility residing in the correlations, out of reach of any symbol-by-symbol code. For the strongly filtered processes below this gap is large: the marginal entropy is close to that of white noise, while the rate is substantially smaller.

Computing \(\bar H\) accurately is difficult. The definition requires the log-probabilities of individual quantization cells, each a high-dimensional Gaussian integral over a box. When the spectral density is close to zero at some frequencies, the covariance matrix is nearly singular and these integrals are hard to evaluate accurately. Such spectra are common in practice: they arise from Gaussian smoothing, from lowpass, highpass, and bandpass filtering, and even from a simple moving average.

Here we present two methods for computing \(\bar H\). The first is an analytical approximation. An exact dithering identity converts the discrete entropy into a Riemann sum for the differential entropy of the dithered process \(X + U\), whose spectral density is exactly \(S(\omega) + \Delta^2/12\); applying the Kolmogorov--Szegő formula then yields
\[\bar H \;\approx\; \frac{1}{4\pi}\int_{-\pi}^{\pi}\log_2\frac{2\pi e\,\big(S(\omega)+\Delta^2/12\big)}{\Delta^2}\,d\omega.\]
This is the classical high-resolution formula, except with an added term representing the noise power of the quantizer. The modified formula is finite for any spectral density, since its integrand is bounded below. Whether it is also accurate is not settled by the derivation, and we address the question empirically in Section 5, using the Monte Carlo estimator described below as ground truth. Across the process families studied there, including strongly filtered processes for which the classical formula fails, the approximation agrees with the exact rate to within a few millibits per sample over most of the parameter range for quantization steps up to about the standard deviation of the signal, the discrepancy growing to a few percent of the rate at the most strongly filtered points; for substantially coarser steps the approximation breaks down and the estimator should be used instead.

The second is a Monte Carlo estimator of the exact rate. Writing \(X_t\) as a finite moving average of i.i.d. Gaussian innovations using its minimum-phase spectral factor makes \(\{Y_t\}\) a hidden Markov process for which the optimal (fully adapted) particle filter is available in closed form. Because the model is conditionally Gaussian with an interval-valued observation, the filter's weights and state updates are both exact, so no proposal approximation enters and the estimate is limited only by the particle count and sequence length. Running the filter on a single long typical sequence and invoking the Shannon--McMillan--Breiman theorem gives an estimate of \(\bar H\) that converges from above as the particle count and sequence length grow together. The estimator may be used directly, or to assess the accuracy of the analytical approximation.

The individual ingredients (high-resolution quantization theory \cite{GrayNeuhoff1998}, dithered quantization \cite{GrayStockham1993,ZamirFeder1996}, the Kolmogorov--Szegő formula \cite{GrenanderSzego1958}, and particle filtering \cite{PittShephard1999}) are all well established. The closest prior work on the approximation is that of Zamir and Feder \cite{ZamirFeder1996}, who analyzed the information rates of dithered (lattice) quantizers applied to filtered Gaussian sources; the noise power \(\Delta^2/12\) enters their analysis in essentially the role it plays in the formula above. Their subject, however, is a genuinely dithered system, characterized in rate--distortion terms, whereas here the dither is purely an analytical device, introduced through the exact identity of Section 3.1, and the quantity approximated is the entropy rate of the undithered quantizer output at finite \(\Delta\). Geiger and Koch \cite{GeigerKoch2019} studied the entropy rate of the same undithered process in the fine-quantization limit: normalizing by \(\log(1/\Delta)\) and letting \(\Delta \to 0\), they showed that the limit, which they call the information dimension rate, equals the fraction of the band on which the spectral density is positive. Their proof likewise employs a dithering identity and the noise floor \(\Delta^2/12\), but the normalization retains only the leading term.

Tamir \cite{Tamir2023} proved a rigorous upper bound of the same spectral form for any stationary integer-valued process, \(\tfrac12\log_2(2\pi e) + \tfrac{1}{4\pi}\int_{-\pi}^{\pi}\log_2\big(\Phi_Y(\omega) + \tfrac{1}{12}\big)\,d\omega\), by combining a dither identity with the Kolmogorov--Szegő formula, and applied it to quantized Gaussian moving-average and autoregressive processes. That integrand uses the spectral density \(\Phi_Y\) of the quantized output, which must be computed separately for each process, whereas the formula above uses the input spectrum \(S\) directly; and because the dithered output there carries the quantization error in addition to the dither, its noise floor is roughly twice the \(\Delta^2/12\) that enters the object \(X + U\) used here, so the bound is looser by that doubled term. Tamir's result is a rigorous bound whereas ours is only an approximation, but our approximation is much closer to the true rate: across most of the experiments of Section 5 it lies within a few millibits of the exact rate, well inside Tamir's bound. To our knowledge the approximation formula above, as a finite-\(\Delta\) description of the undithered quantizer in terms of the input spectrum, has not previously appeared.

On the computational side, the reduction of an information rate to the log-likelihood of a filter evaluated on a single long typical sequence is well established for channels with memory: Arnold et al.\ \cite{ArnoldEtAl2006} developed the approach for finite-state channels using the exact forward recursion, and Dauwels and Loeliger \cite{DauwelsLoeliger2008} extended it to continuous-state models by particle methods. Note that this is a different problem from estimating an entropy rate from observed data, for which plug-in, compression-based, and model-based estimators are reviewed by Feutrill and Roughan \cite{FeutrillRoughan2021}; here the process is specified by its spectral density, and the task is to compute its rate. The contribution of Section 4 is the application of this construction to the uniformly quantized stationary Gaussian process, for which the minimum-phase moving-average representation makes the optimal (fully adapted) particle filter available in closed form.

The paper is organized as follows. Section 2 reviews the classical high-resolution formula and describes the situations in which it fails. Section 3 derives the analytical approximation. Section 4 presents the particle-filter estimator together with its diagnostics. Section 5 reports numerical experiments, and Section 6 concludes with a discussion of limitations and future work; the accompanying implementations are described in Appendix D.

\section{The classical high-resolution formula and its failure}\label{the-classical-high-resolution-formula-and-its-failure}

For small \(\Delta\), the classical approach assumes that the density \(f\) of \(X_1^n\) is approximately constant over each quantization cell \cite{GrayNeuhoff1998}, so that
\[P(Y_1^n = y) = \int_{\prod_t \mathcal C_{y_t}} f(x)\,dx \;\approx\; \Delta^n f(\Delta y).\]
Substituting into the entropy sum and recognizing a Riemann sum with spacing \(\Delta\) \cite[Ch.~8]{CoverThomas2006},
\[H(Y_1,\dots,Y_n) \;\approx\; -\sum_{y\in\mathbb Z^n}\Delta^n f(\Delta y)\,\log_2 f(\Delta y) - n\log_2\Delta \;\approx\; h(X_1^n) - n\log_2\Delta,\]
where \(h\) denotes differential entropy in bits. Since \(X\) is exactly Gaussian, the Kolmogorov--Szegő formula gives its differential entropy rate directly, and dividing by \(n\) and letting \(n\to\infty\) we obtain

\[\bar H_{\mathrm{classical}} \;=\; \tfrac12\log_2(2\pi e) + \frac{1}{4\pi}\int_{-\pi}^{\pi}\log_2 S(\omega)\,d\omega - \log_2\Delta \;=\; \frac{1}{4\pi}\int_{-\pi}^{\pi}\log_2\frac{2\pi e\,S(\omega)}{\Delta^2}\,d\omega.\]

The problem with this formula is apparent: the integrand contains \(\log_2 S(\omega)\), which diverges to \(-\infty\) wherever \(S(\omega)\to 0\). If the spectral density vanishes on part of the frequency axis, the formula predicts an arbitrarily negative entropy rate, impossible since \(H(Y_1^n) \ge 0\). The divergence is the clearest sign of trouble, but the real failure is in the constant-density assumption, and it sets in earlier: wherever the spectral power is small, some linear combination of the samples has standard deviation far below \(\Delta\), so \(f\) is far from constant across a cell in that direction. The approximation therefore degrades gradually, well before \(S\) reaches zero anywhere.

As noted in the introduction, spectra of this kind arise readily in practice. Any linear operation that strongly attenuates part of the spectrum, including Gaussian smoothing, lowpass, highpass, and bandpass filtering, and even a simple moving average, drives \(S\) toward zero on a set of frequencies and puts the classical formula out of reach.

The fine-quantization behavior in this regime is known. Geiger and Koch \cite{GeigerKoch2019} showed that \(\bar H/\log_2(1/\Delta)\) converges, as \(\Delta \to 0\), to the fraction of the band on which \(S\) is positive, a quantity they call the information dimension rate of the process. What is missing is a description at finite \(\Delta\), which is the regime of any real measured signal, and this is the subject of the next section.

\section{The analytical approximation}\label{the-analytical-approximation}

The derivation follows the same outline as in Section 2, except that the cell probabilities are expressed exactly rather than approximately, by way of an identity relating them to the density of a dithered version of the process.

\subsection{The dither identity}\label{the-dither-identity}

Let \(U_t \sim \mathrm{Unif}(-\Delta/2,\Delta/2)\), i.i.d. and independent of \(X\). With \(f\) the density of \(X_1^n\), the density \(g\) of \(X_1^n + U_1^n\) is
\[g(u) = \Delta^{-n}\int_{\prod_t [u_t - \Delta/2,\; u_t + \Delta/2]} f(x)\,dx.\]
Substituting \(u = \Delta y\) for \(y\in\mathbb Z^n\) gives
\[P(Y_1^n = y) = \Delta^n\,g(\Delta y).\]
In other words, the probability of observing the quantized output \(y\) is exactly equal to the density of the dithered process \(X + U\) evaluated at the lattice point \(\Delta y\), multiplied by the cell volume \(\Delta^n\); identities of this kind are standard in the analysis of dithered quantization \cite{GrayStockham1993}.

\subsection{The Riemann-sum step and its lattice error}\label{the-riemann-sum-step-and-its-lattice-error}

The differential entropy of \(g\) is
\[h(X_1^n + U_1^n) = -\int_{\mathbb R^n} g(u)\,\log_2 g(u)\,du,\]
which is approximately equal to its Riemann sum with spacing \(\Delta\):
\[h(X_1^n + U_1^n) \approx -\sum_{y\in\mathbb Z^n}\Delta^n\,g(\Delta y)\,\log_2 g(\Delta y).\]
Substituting \(\Delta^n g(\Delta y) = P(Y_1^n = y)\), the right-hand side equals \(H(Y_1^n) + n\log_2\Delta\), so that
\[H(Y_1,\dots,Y_n) \;\approx\; h(X_1^n + U_1^n) - n\log_2\Delta.\]
Writing \(F(u) = -g(u)\log_2 g(u)\), the Poisson summation formula gives
\[\varepsilon_{\mathrm{lat}}^{(n)} \;:=\; h(X_1^n + U_1^n) - n\log_2\Delta - H(Y_1,\dots,Y_n) \;=\; -\sum_{k\in\mathbb Z^n\setminus\{0\}}\hat F(2\pi k/\Delta), \qquad \hat F(\xi) = \int_{\mathbb R^n} F(u)\,e^{-i\xi\cdot u}\,du.\]
The error is thus a sum of Fourier coefficients of \(F\) evaluated at the nonzero reciprocal lattice points \(2\pi k/\Delta\), which recede to infinity as \(\Delta\to 0\). Its magnitude depends on the rate of decay of \(\hat F\), that is, on the smoothness of \(g\log_2 g\). If \(F\) has \(p\) integrable derivatives then the error is \(O(\Delta^p)\), and for smooth \(F\) it decays faster than any power of \(\Delta\). We do not pursue this analysis here.

\subsection{The spectrum of the dithered process and the Kolmogorov--Szegő step}\label{the-spectrum-of-the-dithered-process-and-the-kolmogorovszegux151-step}

\(U_t\) is i.i.d. with mean zero and variance \(\frac1\Delta\int_{-\Delta/2}^{\Delta/2}u^2\,du = \Delta^2/12 =: v\), so its spectral density is the constant \(v\). Since \(U\) is independent of \(X\), the spectral densities add, and \(X + U\) has spectral density exactly \(S(\omega) + v\).

We approximate \(X + U\) by the Gaussian process with the same spectral density. By the Kolmogorov--Szegő formula \cite{GrenanderSzego1958}, a stationary Gaussian process with spectral density \(S_d\) has differential entropy rate

\[\tfrac12\log_2(2\pi e) + \frac{1}{4\pi}\int_{-\pi}^{\pi}\log_2 S_d(\omega)\,d\omega.\]

Since the Gaussian maximizes entropy for a given covariance \cite{CoverThomas2006}, this is an upper bound on the differential entropy rate of \(X + U\).

Combining these steps, we divide \(H(Y_1^n) \approx h(X_1^n + U_1^n) - n\log_2\Delta\) by \(n\), let \(n\to\infty\), and substitute \(S_d = S + v\):
\[\boxed{\;\bar H \;\approx\; \frac{1}{4\pi}\int_{-\pi}^{\pi}\log_2\frac{2\pi e\,\big(S(\omega)+\Delta^2/12\big)}{\Delta^2}\,d\omega.\;}\]
This is the classical formula of Section 2 with \(S\) replaced by \(S + v\), and the replacement repairs the failure identified there. Since the spectrum is bounded below by \(v = \Delta^2/12\), the integrand saturates at \(\log_2(2\pi e\,v/\Delta^2) = \log_2(2\pi e/12)\) at frequencies where \(S \ll v\), rather than diverging.

Note that the formula also recovers the fine-quantization limit of Geiger and Koch \cite{GeigerKoch2019} described in Section 2: as \(\Delta \to 0\) the integrand grows like \(2\log_2(1/\Delta)\) at frequencies where \(S > 0\) and tends to the constant \(\log_2(2\pi e/12)\) where \(S = 0\), so that the right-hand side divided by \(\log_2(1/\Delta)\) converges to the information dimension rate. The approximation is thus finite in the situations where the classical formula fails, and its leading behavior there is correct.

\subsection{Sources of error}\label{sources-of-error}

Only two approximations were made in the derivation; the remaining steps are exact.

The first is the lattice error \(\varepsilon_{\mathrm{lat}}^{(n)}\) from the Riemann-sum step, given exactly by the Poisson sum above. It is small to the extent that the Fourier coefficients of \(g\log_2 g\) decay rapidly at the reciprocal lattice points; we have not attempted to quantify it here.

The second is the error of the Gaussian approximation: \(X + U\) is not Gaussian (\(U\) is uniform), and the Kolmogorov--Szegő rate for the spectral density \(S + v\) is an upper bound on the true differential entropy rate. The slack in this bound is the per-sample non-Gaussianity of \(X + U\), which is presumably small when \(S \gg v\) over most of the band, since the uniform component is then a mild perturbation of the Gaussian. Note that this error always acts in the same direction: the formula overestimates \(\bar H\).

\subsection{Domain of validity}\label{domain-of-validity}

The approximation appears to remain accurate well into the coarse-quantization regime, failing only when \(\Delta\) is much larger than the scale of the signal. In that extreme regime the formula tends to the constant floor \(\tfrac12\log_2(2\pi e/12)\) while the true rate tends to zero, an artifact of modeling rounding as additive noise. The value of the floor, approximately \(0.25\) bits, is a familiar constant of high-resolution quantization theory. For a Gaussian source, uniformly quantizing at mean-squared distortion \(D\) and entropy-coding the result achieves a rate of \(R(D) + \tfrac12\log_2(2\pi e/12)\), where \(R(D)\) is the rate--distortion function; the excess is the same constant, the asymptotic penalty of uniform quantization over the optimum \cite{GrayNeuhoff1998,GishPierce1968}. Such extreme quantization is not typical of applications. In the electrophysiology recordings that motivate this work the noise standard deviation is at least about twice the quantization step, comfortably within the range of validity. Section 5 maps the accuracy empirically, using the estimator of Section 4 as ground truth.

\section{The particle-filter estimator}\label{the-particle-filter-estimator}

The approximation of Section 3 involves two error terms that have not been quantified, and an independent method for computing the exact rate is therefore needed, both for direct use and for assessing the accuracy of the approximation.

\subsection{Reduction to one typical sequence}\label{reduction-to-one-typical-sequence}

Since \(\{Y_t\}\) is stationary and ergodic (it is a pointwise function of an ergodic Gaussian process, and ergodicity holds provided \(S\) has no atoms), the Shannon--McMillan--Breiman theorem \cite{CoverThomas2006,Chung1961} gives
\[\bar H = \lim_{n\to\infty} -\tfrac1n \log_2 P(Y_1^n) \quad \text{a.s.}\]
Note that the alphabet here is countably infinite, so the usual finite-alphabet statement of the theorem does not literally apply; the extension to countable alphabets is due to Chung \cite{Chung1961}, and its hypothesis \(H(Y_1) < \infty\) is satisfied because \(X_t\) has finite variance, as noted in Section 1. Thus the problem reduces to computing the log-likelihood of a single long typical sequence \(y_1^n\) drawn from the process itself.

\subsection{Moving-average model and hidden Markov structure}\label{moving-average-model-and-hidden-markov-structure}

We represent \(X_t\) as a finite moving average of i.i.d. standard Gaussian innovations:
\[X_t = \sum_{j=0}^{L-1} h_j\,W_{t-j}, \qquad W_t \overset{\text{iid}}{\sim}\mathcal N(0,1),\]
with taps \(h\) chosen so that the model autocovariance \(\gamma^h_k = \sum_j h_j h_{j+k}\) matches the target \(\gamma_k\) (equivalently \(|\hat h(\omega)|^2 = S(\omega)\)). When \(S\) is the squared magnitude of a finite kernel the minimum-phase factor has that same length, so \(L\) is exactly the kernel length; when the factor is infinite, as for an autoregressive spectrum, \(L\) is truncated generously and the residual \(\max_k|\gamma^h_k - \gamma_k|\) confirms the fit. Among the possible spectral square roots we choose the minimum-phase factor, which is computed using the standard cepstral construction \cite[Ch.~13]{OppenheimSchafer2010} (see the implementation of Appendix D). This choice is important for the filtering application since it maximizes the leading tap, giving
\[h_0^2 = \exp\!\Big(\tfrac{1}{2\pi}\int_{-\pi}^{\pi}\ln S(\omega)\,d\omega\Big),\]
the one-step prediction error variance \cite{GrenanderSzego1958}. We assume throughout that \(\int_{-\pi}^{\pi}\ln S(\omega)\,d\omega > -\infty\), so that \(h_0 > 0\); a process whose spectral density vanishes on a set of positive measure is perfectly predictable from its past and admits no such representation. Since \(h_0\) is the conditional standard deviation of \(X_t\) given the past innovations, a symmetric (zero-phase) square root, for which the leading tap is very small, would render the filter degenerate.

Because \(X_t\) depends only on \((W_{t-L+1},\dots,W_t)\), the process \(\{Y_t\}\) is a hidden Markov process with continuous state \(u_t = (W_{t-L+2},\dots,W_t)\in\mathbb R^{L-1}\). The likelihood factors as \(\log_2 P(y_1^n) = \sum_t \log_2 P(y_t\mid y_1^{t-1})\), and each predictive probability is estimated using a particle filter.

\subsection{The fully adapted filter}\label{the-fully-adapted-filter}

The structure of the model admits the optimal (fully adapted) particle filter \cite{PittShephard1999} in closed form. Given the past innovations \(u\), we have \(X_t = h_0 W_t + \mu\) with \(\mu = \sum_{j=1}^{L-1} h_j u_{t-j}\) and \(W_t \sim \mathcal N(0,1)\) independent of the past. Writing \(\Phi\) for the standard normal CDF and expressing the observed cell in standardized units,
\[l = \frac{(y_t-\tfrac12)\Delta - \mu}{h_0}, \qquad r = \frac{(y_t+\tfrac12)\Delta - \mu}{h_0}, \qquad r - l = \frac{\Delta}{h_0},\]
we see that the observation probability is a Gaussian cell probability, \(P(Y_t = y_t \mid u) = \Phi(r) - \Phi(l)\), and that conditional on \(\{Y_t = y_t\}\) and \(u\), the new innovation \(W_t\) is \(\mathcal N(0,1)\) truncated to \([l, r)\).

Note that the dimensionless ratio \(\Delta/h_0\), the quantization step divided by the one-step prediction error, is the key parameter: \(\Delta/h_0 \gg 1\) corresponds to coarse quantization and \(\Delta/h_0 \ll 1\) to fine quantization.

Each of \(N\) particles stores the last \(L-1\) innovations. Resampling is performed at every step, which keeps the weights uniform, so the predictive estimate is a plain average.

\begin{quote}
\textbf{Algorithm.}

\textbf{Inputs:} taps \(h_0 > 0, h_1, \dots, h_{L-1}\); step \(\Delta\); sequence \(y_1^n \in \mathbb Z^n\); particle count \(N\).

Initialize \(u^{(i)}_{2-L},\dots,u^{(i)}_0 \overset{\text{iid}}{\sim} \mathcal N(0,1)\) for each \(i\) (this is the exact prior, so no burn-in is needed). Set \(\ell \leftarrow 0\) (accumulated in nats).

For \(t = 1,\dots,n\):

\begin{enumerate}
\def\labelenumi{\arabic{enumi}.}
\tightlist
\item
  \textbf{Weights:} for each \(i\): \(\mu^{(i)} = \sum_{j=1}^{L-1} h_j u^{(i)}_{t-j}\); \(\;l^{(i)} = \big((y_t-\tfrac12)\Delta - \mu^{(i)}\big)/h_0\), \(\;r^{(i)} = l^{(i)} + \Delta/h_0\); \(\;\alpha^{(i)} = \Phi(r^{(i)}) - \Phi(l^{(i)})\).
\item
  \textbf{Likelihood increment:} \(\ell \leftarrow \ell + \ln\big(\tfrac1N\sum_i \alpha^{(i)}\big)\).
\item
  \textbf{Resample:} draw ancestor indices by systematic resampling with probabilities \(\propto \alpha^{(i)}\); set each particle's innovation buffer to its ancestor's.
\item
  \textbf{Propagate:} for each \(i\), draw \(u^{(i)}_t \sim \mathcal N(0,1)\) truncated to \([l^{(a_i)}, r^{(a_i)})\).
\end{enumerate}

\textbf{Output:} \(\hat{\bar H} = -\ell/(n\ln 2)\) bits per sample.
\end{quote}

Step 4 samples exactly from \(p(w_t \mid y_t, u)\) and step 3 resamples using the exact predictive weights, so this is the fully adapted auxiliary particle filter for this model. Del Moral's proposition \cite[Prop.~7.4.1]{DelMoral2004} is stated for multinomial resampling, but the estimator \(\prod_t \frac1N\sum_i \alpha^{(i)}\) remains unbiased for \(P(y_1^n)\) under systematic resampling as well, since its offspring counts are conditionally unbiased, and we use systematic resampling here for its lower variance. The computational cost is \(O(nNL)\) time and \(O(NL)\) memory. Numerically stable evaluation of the cell probabilities and of the truncated-normal draws requires some care in log space; the details are given in Appendix A and in the accompanying code.

\subsection{Bias, convergence, and reporting}\label{bias-convergence-and-reporting}

The estimator overestimates \(\bar H\) in expectation for two reasons. First, by Jensen's inequality, \(\mathbb E\log\hat P(y_1^n) \le \log P(y_1^n)\), since the particle likelihood is unbiased on the natural scale but not on the log scale; the resulting bias is of order \(O(1/N)\) per sample; Section 5.1 observes this rate for processes of moderate difficulty (for the hardest processes the decay is slower at practical particle counts). Second, the finite-\(n\) block entropy \(\tfrac1n H(Y_1^n)\) is non-increasing in \(n\) and converges to \(\bar H\) \cite{CoverThomas2006}. Both effects vanish in the appropriate limit, so that \(\hat{\bar H}\) approaches \(\bar H\) from above and the practical convergence check is to increase \(N\) and \(n\) until the estimate stops decreasing. Results are reported as the mean \(\pm\) standard error over independent replicates, each with a freshly generated sequence and filter, and Section 5 includes a convergence study with respect to both \(N\) and \(n\).

For strongly smoothing processes the particle cloud can lose track of the state entirely, in which case the resulting estimate is invalid rather than merely imprecise, and the effective sample size does not detect the failure. The failure mode, together with a diagnostic based on the per-step likelihood increment that does detect it, is described in Appendix B.

\section{Numerical experiments}\label{numerical-experiments}

This section reports numerical experiments over a range of process families (white noise, the first difference, moving averages, AR(1), Gaussian smoothing, and lowpass filtering) and quantization steps. Section 5.1 studies the convergence of the particle-filter estimator in its two parameters and establishes how they should be chosen. Section 5.2 then compares the analytical approximation of Section 3 against the resulting estimates across the process families, and, within the validated range, surveys the dependence of the entropy rate on the process and quantization parameters. Finally, Section 5.3 compares practical lossless coders against the estimated rate.

\subsection{Convergence of the particle filter}\label{convergence-of-the-particle-filter}

The estimator of Section 4 has two convergence parameters, the particle count \(N\) and the sequence length \(n\), and the experiments in this subsection establish how they should be chosen. Seven processes spanning the range of difficulty the filter encounters are used, all quantized at \(\Delta = 0.25\sigma\): moving averages of widths 8, 16, 32, and 64, Gaussian smoothing kernels of widths \(\tau = 1\) and \(\tau = 1.25\), and the first difference. Two strongly smoothing processes, for which the filtering problem is hardest, are used to study the collapse diagnostic: a 33-tap windowed-sinc lowpass filter (cutoff 6 kHz at a sampling rate of 30 kHz) at \(\Delta = 0.25\sigma\), and a Gaussian kernel of width \(\tau = 1.5\) at \(\Delta = 0.5\sigma\). All runs used the WebGPU engine described in Appendix D, and scripts reproducing every figure and table are included in the repository.

Figure 1 shows the paired excess of the estimate at particle count \(N\) over the estimate at \(N = 128000\) on the same observed sequences. The pairing cancels the sequence-to-sequence variation and isolates the filter bias. The excess is positive throughout and decreases with \(N\) in every process, as expected from the Jensen bias discussed in Section 4.4, and for the processes of moderate difficulty the decay is consistent with the expected \(O(1/N)\) rate over more than two decades.

\begin{figure}[t]
\centering
\includegraphics[width=0.7\linewidth]{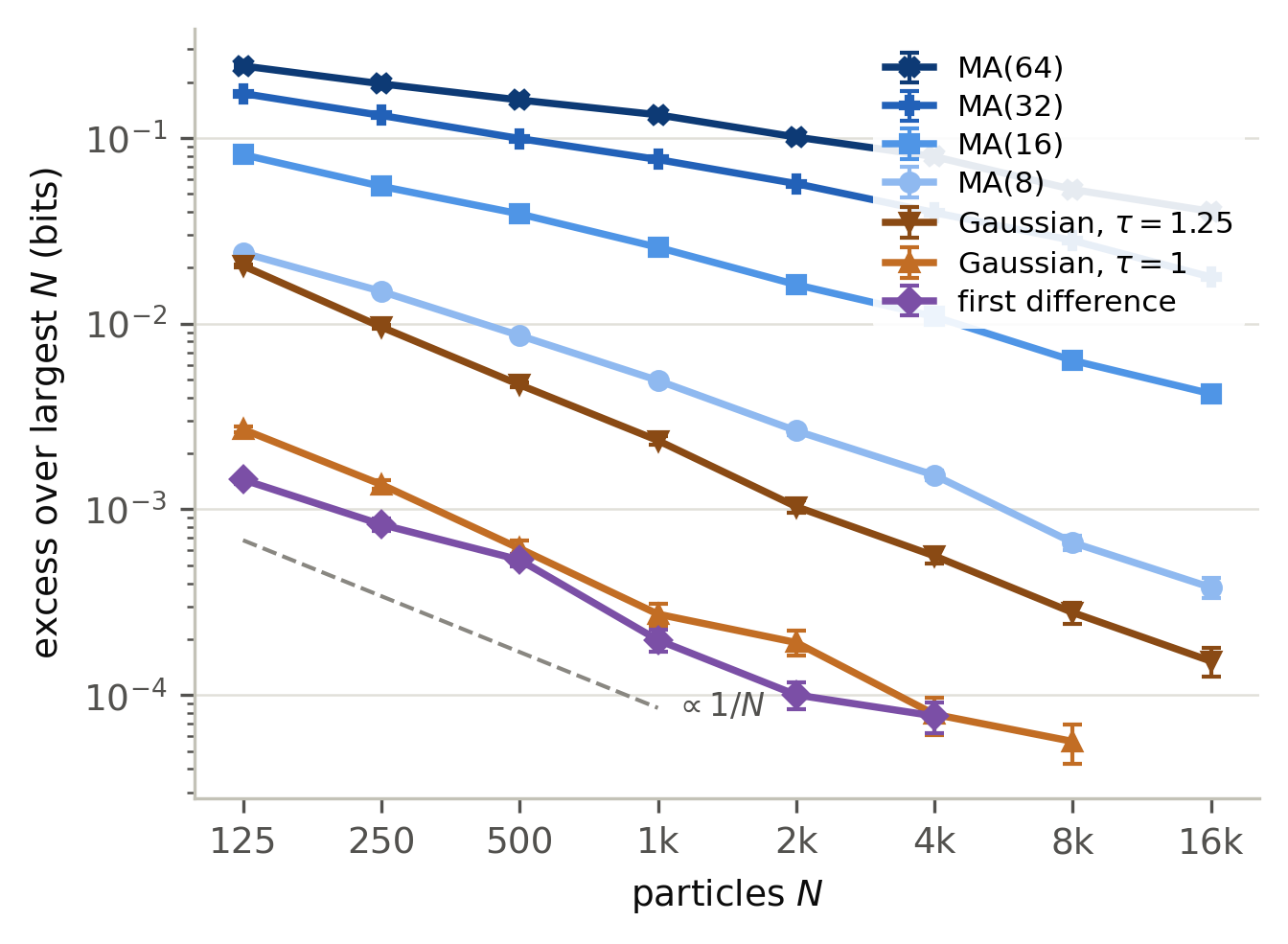}
\caption{Convergence in the particle count $N$: the paired excess of the estimate at $N$ over the estimate at $N = 128000$ on the same observed sequences, at $n = 20000$ with 32 replicates and $\Delta = 0.25\sigma$. Each curve is drawn up to the last $N$ at which the excess is resolved (mean above two standard errors). The dashed guide is proportional to $1/N$; note that the moving-average curves are visibly shallower. Since the reference carries a residual bias of its own, the rightmost points understate the true bias by about 12 percent.}
\label{fig:1}
\end{figure}

The size of the bias at a given \(N\) spans more than two orders of magnitude across the processes, and the variation is evidently not governed by the dimension of the state; the AR(1) process with \(\rho = 0.9\), whose filter carries a 63-dimensional state, is already converged at \(N = 125\) and does not appear in the figure. It is instead related to the structure of the spectrum: the moving averages, whose spectra vanish at the \(w - 1\) nulls of the boxcar response, are the most biased at every width, and for the wider ones the decay is visibly slower than \(1/N\). The \(O(1/N)\) asymptotics presumably still hold, but the particle count at which they set in recedes as the process hardens; the same slow convergence is encountered in Section 5.2. In practice the convergence check of Section 4.4 is inexpensive, and for processes of moderate difficulty a few thousand particles suffice.

Figure 2 shows the corresponding sweep over the sequence length at fixed \(N\). The replicate standard deviation shrinks in proportion to \(1/\sqrt{n}\) for every process across the 64-fold range of lengths, while the means are constant to within their error bars; the finite-\(n\) transient of Section 4.4, which is \(O(L/n)\) of the total, is resolvable only for the widest kernel at the shortest lengths. Thus the role of \(n\) is to increase precision, while the bias is controlled by \(N\) alone. Since the variance of the reported mean scales as the reciprocal of the total sample count \(rn\), a fixed budget may be divided between longer sequences and more replicates as convenience dictates, provided \(n\) is much longer than the correlation length. The experiments of Section 5.2 exploit this freedom: since every step is an opportunity to lose lock, their hardest points are run as many short replicates rather than a few long ones.

\begin{figure}[t]
\centering
\includegraphics[width=0.7\linewidth]{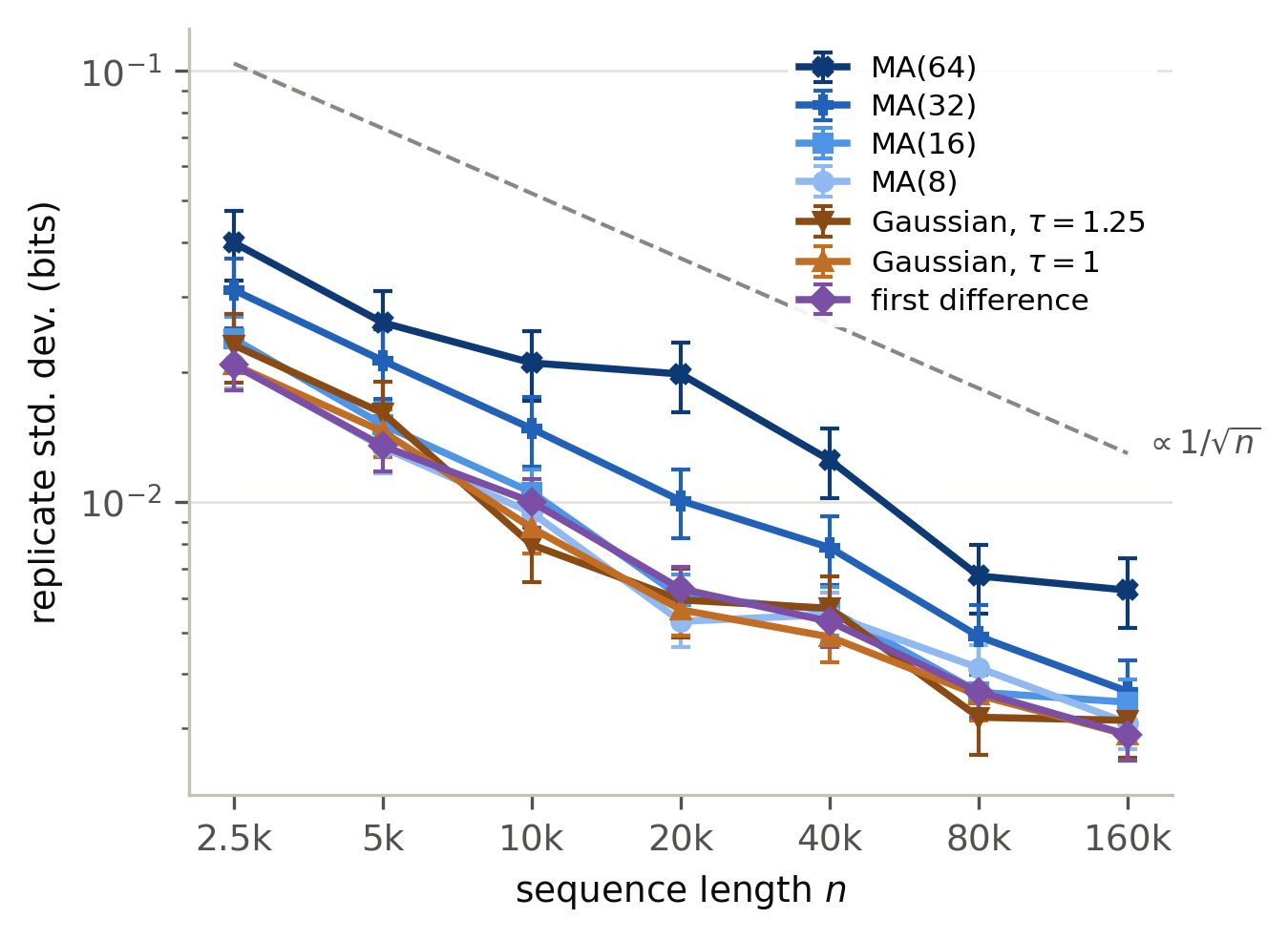}
\caption{Convergence in the sequence length $n$: the standard deviation across replicates at fixed particle count ($N = 4000$; $16000$ for the moving averages of widths 32 and 64), with $\Delta = 0.25\sigma$. Error bars are the standard error of the standard deviation (32 replicates; 16 for the two widest moving averages). The dashed guide is proportional to $1/\sqrt{n}$.}
\label{fig:2}
\end{figure}

For the two strongly smoothing processes the picture changes qualitatively. Figure 3 shows, as a function of the particle count, the fraction of steps flagged by the lost-lock diagnostic of Appendix B together with the mean effective sample size, and Table 1 reports the corresponding estimates. At small \(N\) every replicate loses the state, and the resulting values are invalid, exceeding the memoryless upper bound \(H(Y_1)\) by many orders of magnitude. The transition to reliable operation is fairly sharp for the Gaussian kernel, all replicates keeping lock from \(N = 3000\) onward, whereas the lowpass process, whose state has dimension 32, requires \(N = 100000\). Note that each step is an opportunity to lose the state, so the particle count required grows with the sequence length.

\begin{figure}[t]
\centering
\includegraphics[width=\linewidth]{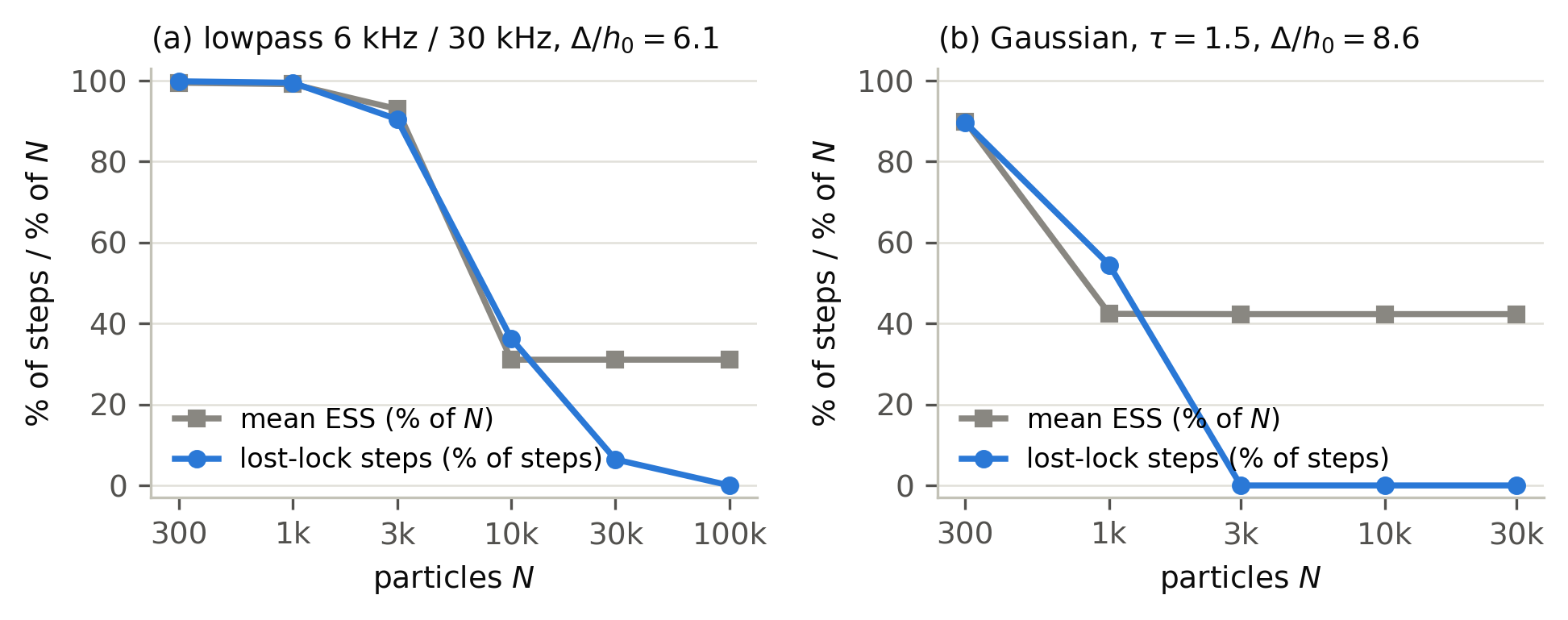}
\caption{Behavior of the collapse diagnostic in the fine-quantization regime, for (a) the lowpass process at $\Delta = 0.25\sigma$ and (b) the Gaussian kernel with $\tau = 1.5$ at $\Delta = 0.5\sigma$, at $n = 30000$ with eight replicates. Shown are the fraction of steps flagged as lost-lock by the diagnostic of Appendix B and the mean effective sample size as a fraction of $N$. Note that the ESS is highest in the collapsed regime.}
\label{fig:3}
\end{figure}

The effective sample size does not detect these failures; it is in fact higher in the collapsed regime than in the healthy one, since once every particle is comparably far from the true state the weights are nearly uniform and the resampling statistics appear ideal. The per-step log-likelihood diagnostic, by contrast, flags the collapsed steps directly, and the affected replicates are discarded rather than averaged in.

\begin{table}[t]
\centering
\begin{tabular}{@{}rrrrr@{}}
\toprule
\(N\) & estimate (bits/sample) & discarded replicates & lost-lock steps & mean ESS \\
\midrule
\multicolumn{5}{@{}l}{\textbf{lowpass 6 kHz / 30 kHz} (\(\Delta = 0.25\sigma\), \(\Delta/h_0 = 6.1\), \(L = 33\), \(n = 30000\))} \\
300 & 7.54e+08 (collapsed) & 8/8 & 99.8\% & 99\% \\
1000 & 7.49e+08 (collapsed) & 8/8 & 99.5\% & 99\% \\
3000 & 5.76e+08 (collapsed) & 8/8 & 90.5\% & 93\% \\
10000 & 2.1874 (provisional) & 5/8 & 36.3\% & 31\% \\
30000 & 2.1817 (provisional) & 2/8 & 6.4\% & 31\% \\
100000 & 2.1806 \(\pm\) 0.0015 & 0/8 & 0.0\% & 31\% \\
\midrule
\multicolumn{5}{@{}l}{\textbf{Gaussian, \(\tau=1.5\)} (\(\Delta = 0.5\sigma\), \(\Delta/h_0 = 8.6\), \(L = 13\), \(n = 30000\))} \\
300 & 2.73e+05 (collapsed) & 8/8 & 89.6\% & 90\% \\
1000 & 1.6403 (provisional) & 7/8 & 54.5\% & 42\% \\
3000 & 1.6380 \(\pm\) 0.0016 & 0/8 & 0.0\% & 42\% \\
10000 & 1.6360 \(\pm\) 0.0018 & 0/8 & 0.0\% & 42\% \\
30000 & 1.6353 \(\pm\) 0.0018 & 0/8 & 0.0\% & 42\% \\
\bottomrule
\end{tabular}
\caption{The collapse study. A run is \emph{collapsed} when every replicate lost the state, and \emph{provisional} when some but not all did, since the surviving replicates were produced by the same particle count that failed elsewhere. Lost-lock steps and mean ESS are averaged over all replicates, including discarded ones.}
\label{tab:collapse}
\end{table}

An estimate for which some replicates were discarded should therefore be treated as provisional, and the working rule is to increase \(N\) until no replicate loses lock. The surviving replicates at the provisional particle counts already agree with the fully locked estimate to within their error bars, so this discipline costs little beyond computation. It is also reassuring that the converged estimates fall slightly below the analytical approximation (\(2.1806 \pm 0.0015\) against \(2.1882\) bits for the lowpass, and \(1.6353 \pm 0.0018\) against \(1.6415\) for the Gaussian kernel), which is the direction predicted by the analysis of Section 3.4.

\subsection{Accuracy of the analytical approximation}\label{accuracy-of-the-analytical-approximation}

The approximation of Section 3 was compared against the particle-filter estimate across six process families: four parameterized families (the moving average as a function of its width \(w\), the AR(1) process as a function of its correlation \(\rho\), the Gaussian kernel as a function of its width \(\tau\), and the windowed-sinc lowpass as a function of its cutoff), shown in Figures 4 through 7, together with white noise and the first difference \([1, -1]\), whose spectrum vanishes at DC, shown in Figure 8. The four family figures share a common format. Panel (a) gives the entropy rate as a function of the family parameter at four quantization steps, \(\Delta = \sigma/8\), \(\sigma/4\), \(\sigma\), and \(2\sigma\), with the approximation drawn on a dense parameter grid and the estimates overlaid at eight parameter values per curve, so that the approximation is checked at every part of each sweep rather than at isolated spot checks. Panel (b) gives the error of the approximation on the millibit scale. In Figure 8 the quantization step itself is on the horizontal axis.

\begin{figure}[t]
\centering
\includegraphics[width=\linewidth]{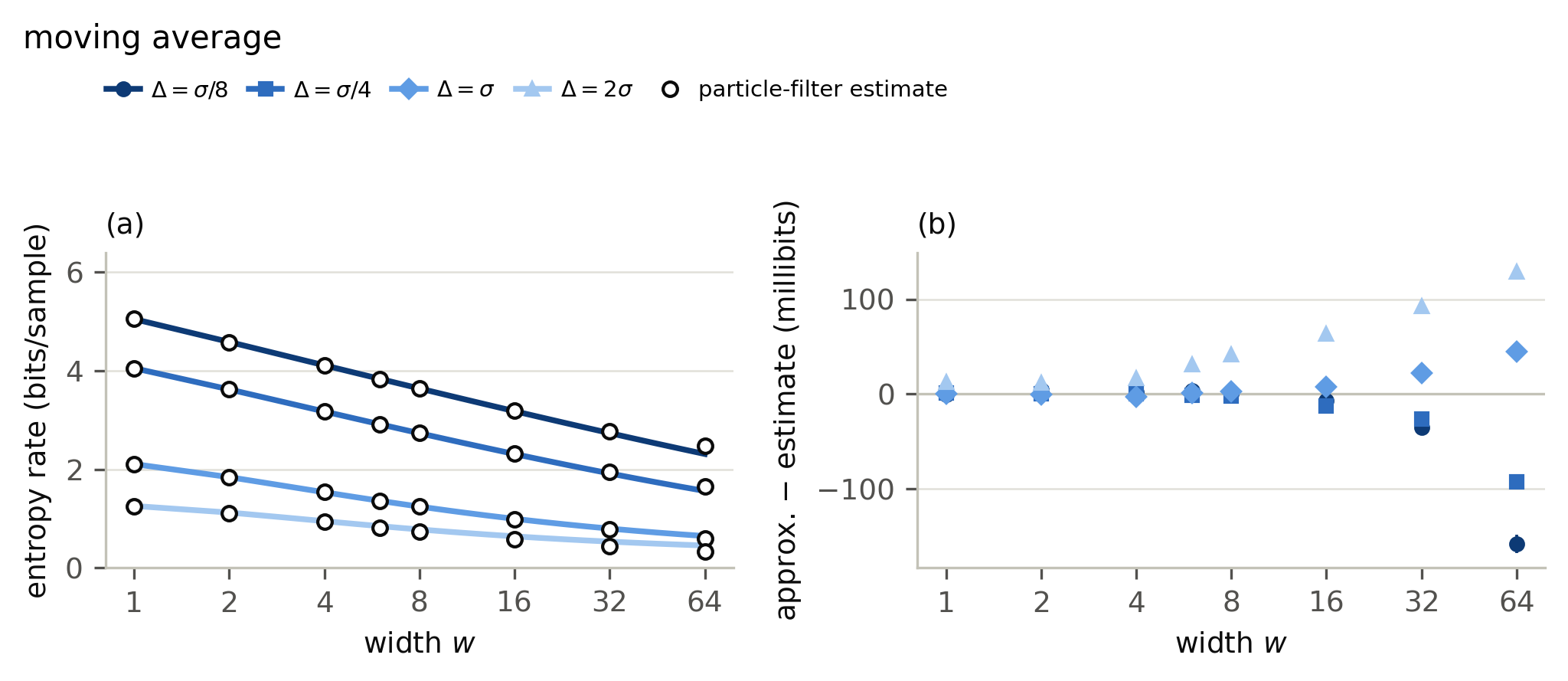}
\caption{The moving average of width $w$ ($\sigma = 1$). (a) The entropy rate: the analytical approximation on a dense grid of widths (curves) at four quantization steps, with the particle-filter estimates overlaid (open circles; error bars are smaller than the markers). (b) The error of the approximation (approximation minus estimate, in millibits per sample) at the measured points, $\pm$ one standard error. The vertical scale of panel (b) differs from one family to the next.}
\label{fig:4}
\end{figure}

\begin{figure}[t]
\centering
\includegraphics[width=\linewidth]{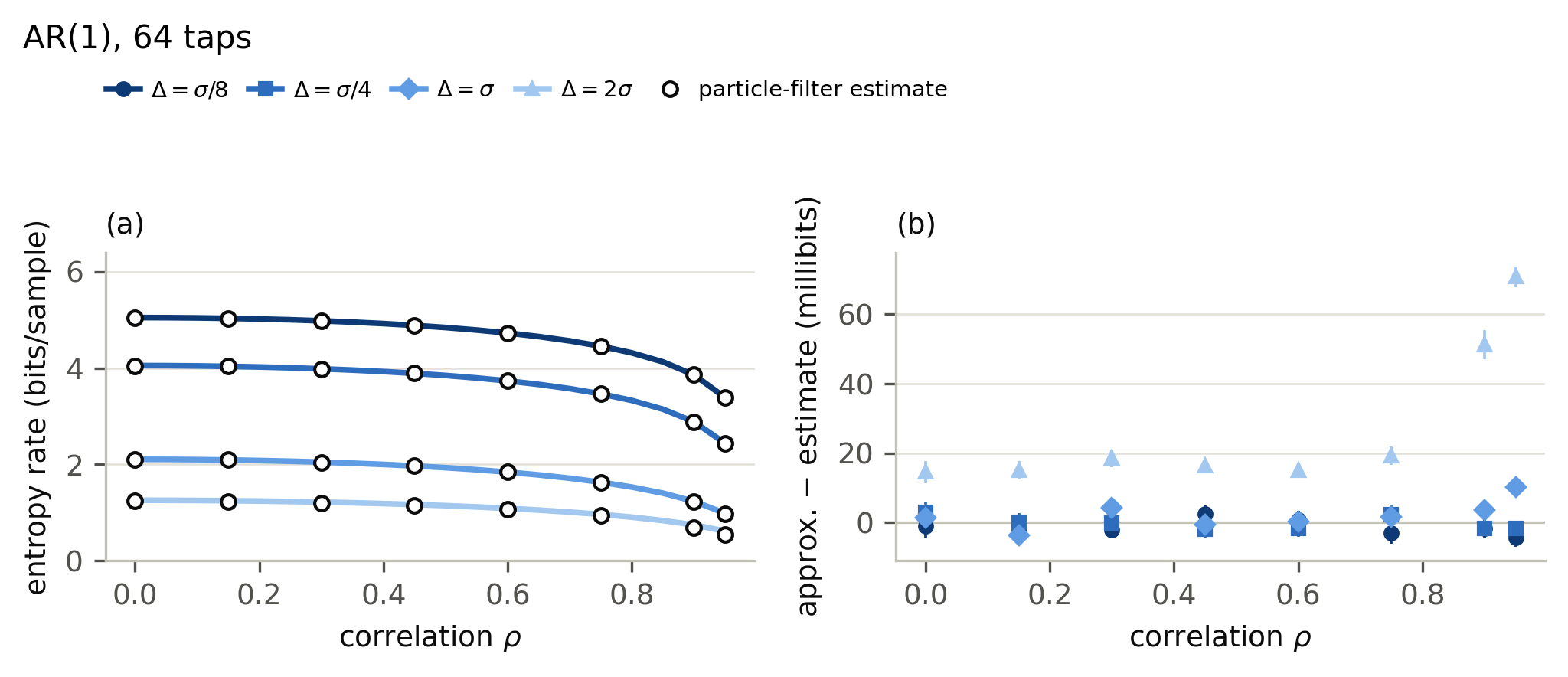}
\caption{The AR(1) process with correlation $\rho$, truncated to 64 taps, in the format of Figure 4.}
\label{fig:5}
\end{figure}

\begin{figure}[t]
\centering
\includegraphics[width=\linewidth]{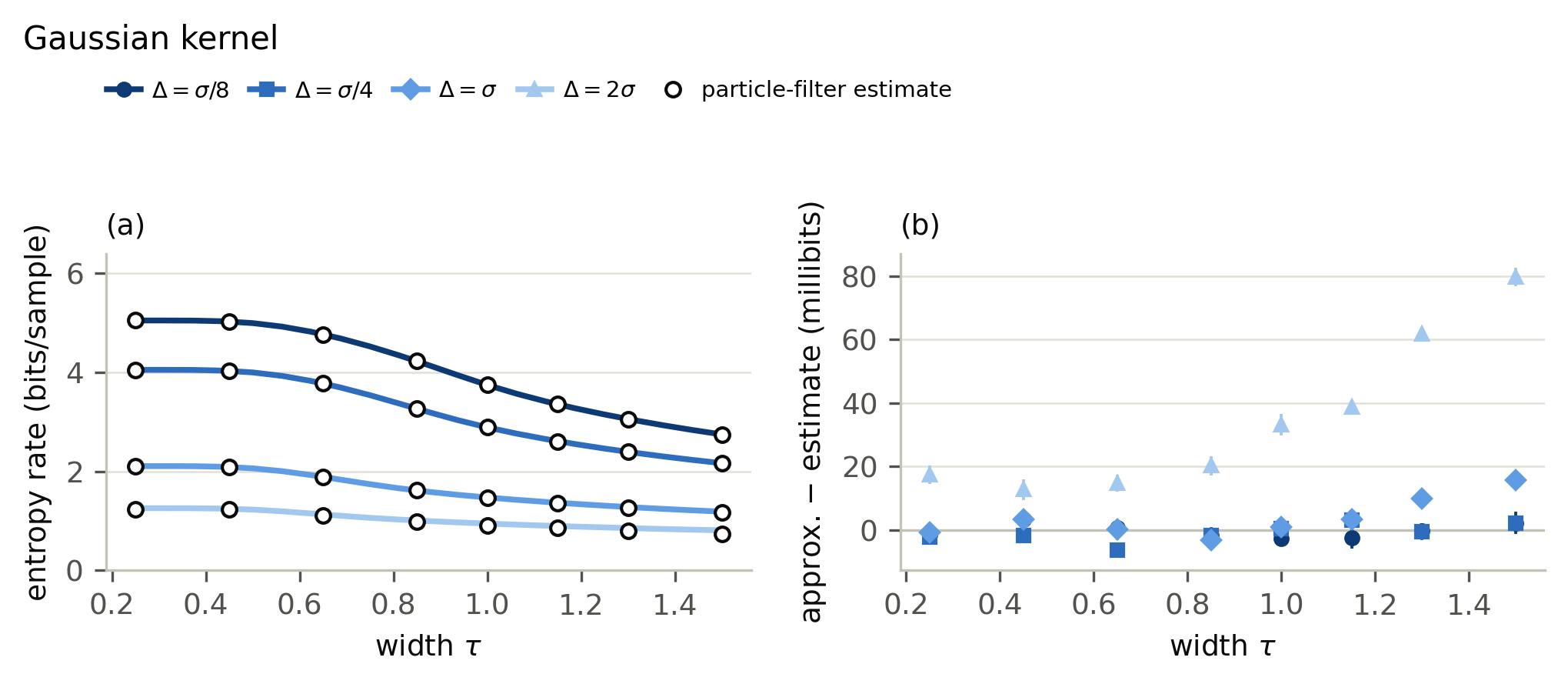}
\caption{The Gaussian kernel of width $\tau$, with $2\lceil 4\tau\rceil + 1$ taps, in the format of Figure 4.}
\label{fig:6}
\end{figure}

\begin{figure}[t]
\centering
\includegraphics[width=\linewidth]{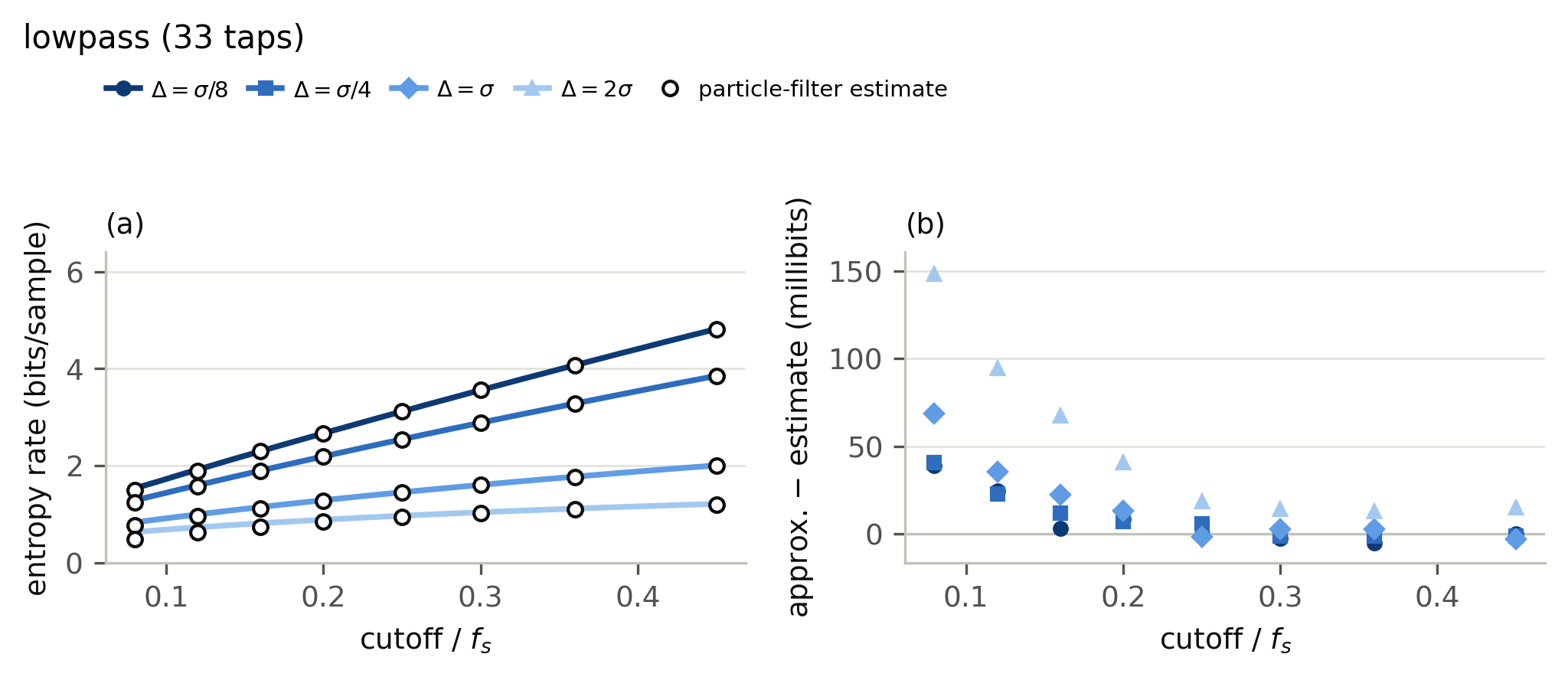}
\caption{The windowed-sinc lowpass filter with 33 taps, as a function of the cutoff frequency, in the format of Figure 4.}
\label{fig:7}
\end{figure}

The filter parameters follow the convergence study of Section 5.1, with the particle count set from the one-step prediction error \(h_0\) of each point, since \(h_0\) governs how hard the filtering problem is: \(N = 8000\) where \(h_0 > 0.3\sigma\), \(N = 30000\) below that, and \(N = 130000\) where \(h_0 < 0.05\sigma\), namely at the lowest four lowpass cutoffs. Sixteen replicates of length \(n = 10000\) are run at every point, reduced to \(n = 5000\) at the hardest ones. Since the particle count needed to keep a replicate locked grows with the sequence length, while the precision of the mean depends only on the total sample count (Section 5.1), the shorter replicates buy the same precision at a much lower particle count; we verified that the shortening introduces no finite-\(n\) bias. Each point uses independently seeded sequences, so the scatter in the error panels is independent from point to point, and the per-point parameters and complete numerical values are recorded in Appendix C. For white noise the symbol probabilities are available in closed form and the likelihood is computed exactly, so that comparison contains no filter error at all.

For \(\Delta \le \sigma\) the estimates fall on the curves throughout, across families, parameter values, and quantization steps, and the rate panels show no visible discrepancy. The error panels resolve what the rate panels cannot. Over most of each sweep the measured error is within a few millibits of zero, comparable to the estimator's own standard error of about 3 millibits and far below any level that matters in practice. Toward the strongly smoothing end of each sweep a systematic overestimate develops, in the direction predicted by the analysis of Section 3.4. It is clearest for the lowpass family, whose error grows from a few millibits at the wider cutoffs to some tens of millibits at the lowest cutoff, roughly five percent of the rate there; restricted to cutoffs of \(0.2 f_s\) and above, the same family agrees to within \(13\) millibits at every step up to \(\sigma\).

At \(\Delta = 2\sigma\) the overestimate becomes systematic in every family and grows with the strength of the smoothing, from 12 to 15 millibits for white noise, the first difference, and the weakly correlated ends of the sweeps, to about 25 percent of the rate for the lowpass at its lowest cutoff. This is the onset of the coarse-quantization failure anticipated in Section 3.5, and it is why the curves in the rate panels visibly separate from the estimates only at this step.

The moving average is a partial exception, and the one place in these experiments where the comparison is limited by the estimator rather than by the formula. Through width 8 it behaves like the other families, agreeing to within a few millibits for \(\Delta \le \sigma\). At the wider kernels and the finer steps, however, the estimate lies systematically \emph{above} the approximation, by more than a hundred millibits at \(w = 64\) and \(\Delta = \sigma/8\). Since both methods are biased upward, a negative measured error of this size cannot be an error of the approximation, and must instead be residual bias of the filter. A separate sweep of the particle count at that point confirms this reading and shows how slow the convergence is, the estimate still decreasing at \(N = 400000\). This is consistent with the finding of Section 5.1 that the moving average is by far the most biased family for the filter, and it means that these particular points neither confirm nor contradict the approximation. Elsewhere in the sweep, and in every other family, the filter is converged and the comparison is meaningful.

Figure 8 carries the comparison for white noise and the first difference out to \(\Delta = 4\sigma\). The estimates track the curves to within about 5 millibits through \(\Delta = \sigma\) and fall below them by 13 to 14 millibits at \(2\sigma\). At \(4\sigma\) the failure is qualitative, the error being about \(0.34\) bits, larger than the rate itself. Indeed, for white noise the approximation can also be checked against the exact marginal entropy, and its true error is below \(0.01\) millibits for \(\Delta \le \sigma\); the differences measured at the fine steps are therefore attributable to the estimator, not the formula, and they calibrate the residual scatter seen in the other families. As \(\Delta \to \infty\) the error approaches the constant floor \(\tfrac12\log_2(2\pi e/12) \approx 0.255\) bits discussed in Section 3.5, and it does so slowly from above, since the residual spectral term of the formula decays only as \(O(\sigma^2/\Delta^2)\) while the true rate vanishes faster.

\begin{figure}[t]
\centering
\includegraphics[width=\linewidth]{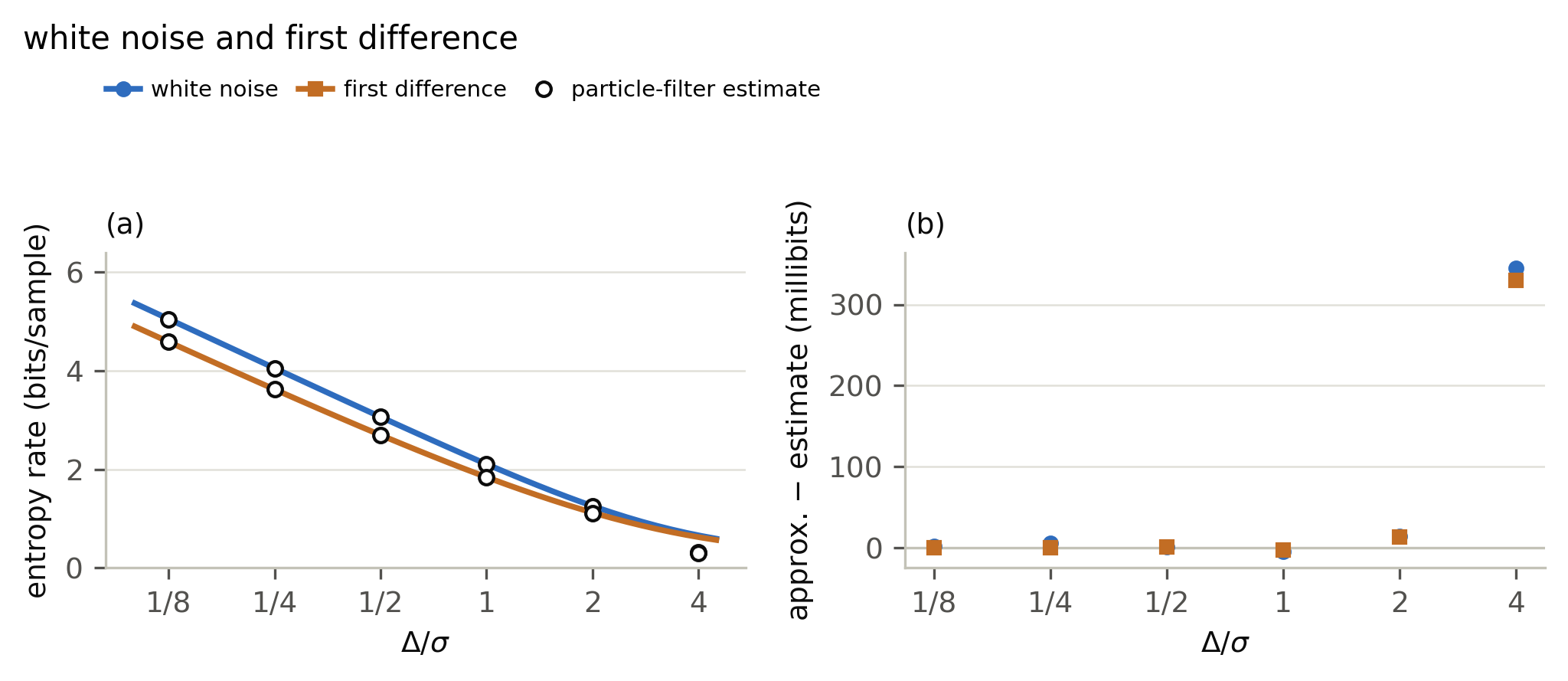}
\caption{White noise and the first difference $[1, -1]$ as functions of the quantization step. (a) The entropy rate: the analytical approximation (curves) with the particle-filter estimates overlaid (open circles); for white noise the likelihood is computed exactly rather than filtered. (b) The error of the approximation at the measured points, $\pm$ one standard error. The rightmost points show the coarse-quantization failure of the approximation.}
\label{fig:8}
\end{figure}

With the accuracy for \(\Delta \le \sigma\) established, the rate panels may also be read as a survey of the dependence of the entropy rate on the process and quantization parameters, mapped densely at negligible cost. In the fine-quantization regime the rate is governed by the one-step prediction error, \(\bar H \approx \tfrac12\log_2(2\pi e\,h_0^2/\Delta^2)\), so the family parameter acts on the rate through \(h_0\). For the moving average \(h_0 = \sigma/\sqrt{w}\), and the top curve of Figure 4(a) falls by very nearly half a bit per doubling of the width; for the lowpass filter the rate is approximately linear in the cutoff, reflecting the bandwidth factor in the Kolmogorov--Szegő integral. The vertical separation between adjacent curves equals \(\log_2\) of the ratio of the steps as long as the quantization is fine relative to the whole spectrum; under strong smoothing part of the spectrum falls below the quantization noise power \(\Delta^2/12\) and the separation compresses, which is visible at the wide end of the moving-average and Gaussian sweeps.

Across the families, the approximation agrees with the estimator to within a few millibits per sample for \(\Delta \lesssim \sigma\), including strongly filtered processes on which the classical formula fails outright, at a cost of milliseconds rather than the minutes to hours of the estimator. The agreement degrades gradually as the smoothing strengthens, reaching a few percent of the rate at the most strongly filtered points tested, and sharply beyond \(\Delta \approx 2\sigma\), where the additive-noise model breaks down. In the regime of interest here (\(\sigma \gtrsim 2\Delta\)) the approximation may replace the estimator for most purposes; beyond \(\Delta \approx 2\sigma\) it should not be used, and the estimator remains available there.

\subsection{Comparison with practical compressors}\label{comparison-with-practical-compressors}

Finally, we compare a panel of lossless coders against the entropy rate, to indicate how close practical compression comes to the ideal limit. Three kinds of coder are included. The first kind is the general-purpose byte compressors, applied to the raw symbol stream serialized as little-endian 16-bit integers, representative of the common practice of running a compressor over the data as stored: zlib\footnote{\url{https://zlib.net}} (level 9), zstd\footnote{\url{https://facebook.github.io/zstd}} (level 19), brotli\footnote{\url{https://github.com/google/brotli}} (quality 11), LZMA (preset 9), and bzip2 (level 9), each at its strongest setting. The second kind models the signal: delta coding, that is, the first difference of the integer stream, followed by zlib or by an ANS entropy coder\footnote{The simple-ans package, \url{https://github.com/flatironinstitute/simple_ans}}; the reference FLAC encoder \cite{RFC9639} at level 8, the standard for lossless audio, which fits its own linear predictor blockwise and Rice-codes the residuals; and a predictive pipeline of our own, the fixed-point linear predictor of the accompanying package (FLAC style, order 32 for every family) followed by ANS on the residuals. The third kind is ANS alone on the raw stream, a coder with no model at all, which realizes the order-0 entropy and serves as a reference point. Each transform is integer-reversible, so every entry is a lossless code of the same data. The library versions were zlib 1.3.1, zstd 1.5.7 (through python-zstandard 0.25.0), Brotli 1.2.0, liblzma 5.8.1, bzip2 1.0.8, libFLAC 1.4.3 (through pyFLAC 3.0.0), and simple-ans 0.3.1, all driven from Python 3.13.5.

Each configuration compresses five independent realizations of length \(n = 10^6\). The reference rate is the analytical approximation, which Section 5.2 validated over this range of steps. Figure 9 follows three representative families as the step varies, Table 2 reports the whole panel at \(\Delta = \sigma/4\), and Figure 10 shows each coder's overhead above the rate, family by family, at \(\Delta = \sigma/4\) and \(\Delta = \sigma\).

\begin{figure}[t]
\centering
\includegraphics[width=\linewidth]{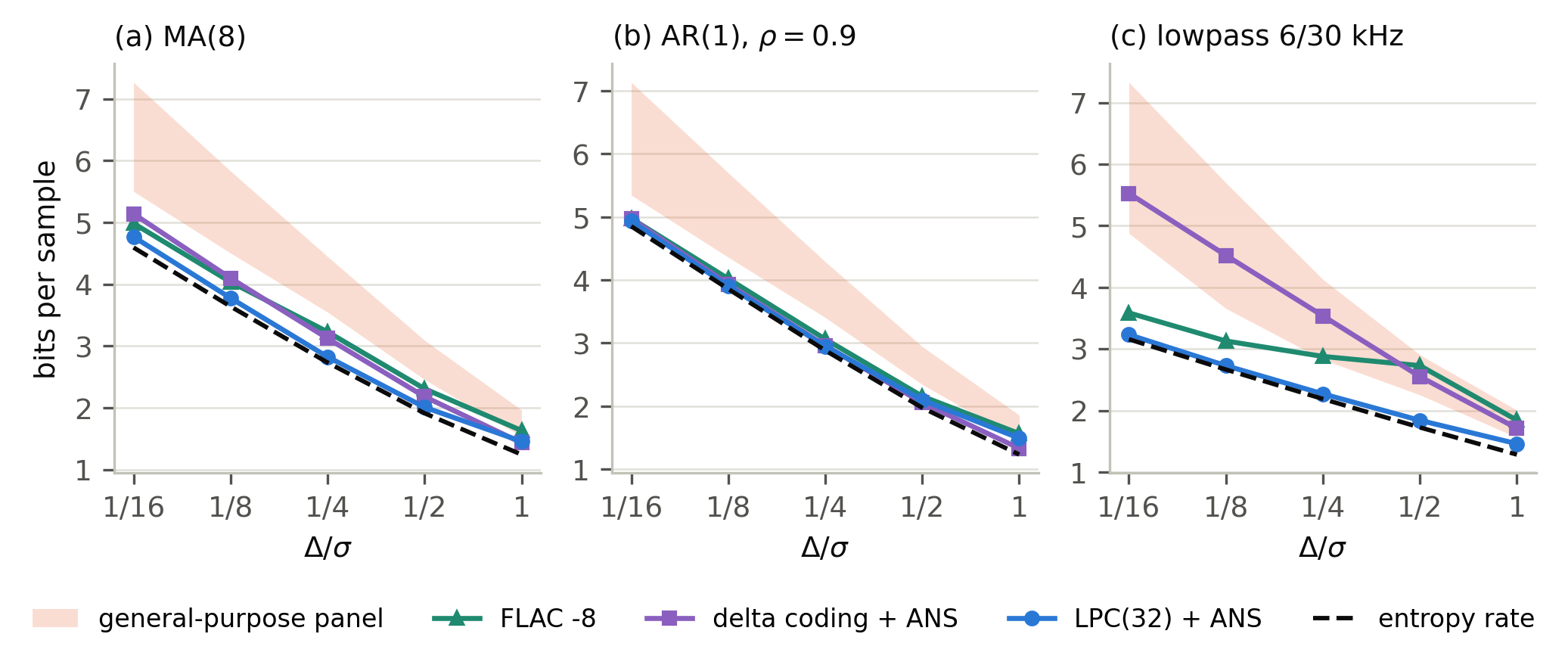}
\caption{Practical coders against the entropy rate (dashed), for (a) the MA(8) process, (b) the AR(1) process, and (c) the lowpass process, as a function of the quantization step. The shaded band spans the five general-purpose compressors on the raw stream. Each point is the mean over five realizations of length $n = 10^6$. For the three coders drawn individually the spread across realizations is at most 0.06 bits per sample; the general-purpose compressors vary more, by up to 0.17 bits, which the edges of the band inherit.}
\label{fig:9}
\end{figure}

\begin{table}[t]
\centering
{\small
\begin{tabular}{@{}lrrrrrrrr@{}}
\toprule
coder & white & diff & MA(8) & AR(1) & $\tau{=}1$ & $\tau{=}1.5$ & lowpass & overhead \\
\midrule
entropy rate & 4.051 & 3.621 & 2.731 & 2.884 & 2.888 & 2.165 & 2.188 & -- \\
order-0 entropy & 4.049 & 4.049 & 4.048 & 4.047 & 4.049 & 4.049 & 4.049 & 45\% \\
\midrule
ANS on raw & 4.095 & 4.093 & 4.090 & 4.090 & 4.093 & 4.091 & 4.093 & 46\% \\
\midrule
zlib -9 & 5.484 & 5.229 & 4.449 & 4.291 & 4.564 & 3.745 & 4.130 & 59\% \\
zstd -19 & 4.963 & 4.641 & 3.838 & 3.655 & 3.901 & 3.037 & 3.334 & 35\% \\
brotli -11 & 4.634 & 4.453 & 3.746 & 3.597 & 3.779 & 3.003 & 3.275 & 31\% \\
LZMA -9 & 4.537 & 4.257 & 3.547 & 3.405 & 3.602 & 2.869 & 3.136 & 25\% \\
bzip2 -9 & 4.618 & 4.299 & 3.549 & 3.409 & 3.461 & 2.659 & 2.827 & 22\% \\
\midrule
delta coding + zlib -9 & 5.891 & 6.019 & 4.202 & 3.991 & 4.353 & 3.340 & 3.800 & 55\% \\
delta coding + ANS & 4.612 & 4.914 & 3.122 & 2.961 & 3.539 & 2.995 & 3.537 & 27\% \\
FLAC -8 & 4.164 & 3.885 & 3.228 & 3.062 & 3.234 & 2.847 & 2.879 & 16\% \\
LPC(32) + ANS & 4.114 & 3.687 & 2.824 & 2.952 & 2.957 & 2.242 & 2.271 & 3\% \\
\bottomrule
\end{tabular}
}
\caption{Bits per sample achieved by every coder at \(\Delta = \sigma/4\), against the entropy rate and the order-0 (memoryless) symbol entropy, with the mean overhead relative to the entropy rate, averaged over the seven families, in the last column. Within each block the coders are ordered by that mean. The order-0 entropy is nearly the same for every family, since the marginal distribution does not depend on the correlation structure.}
\label{tab:compressors}
\end{table}

\begin{figure}[t]
\centering
\includegraphics[width=\linewidth]{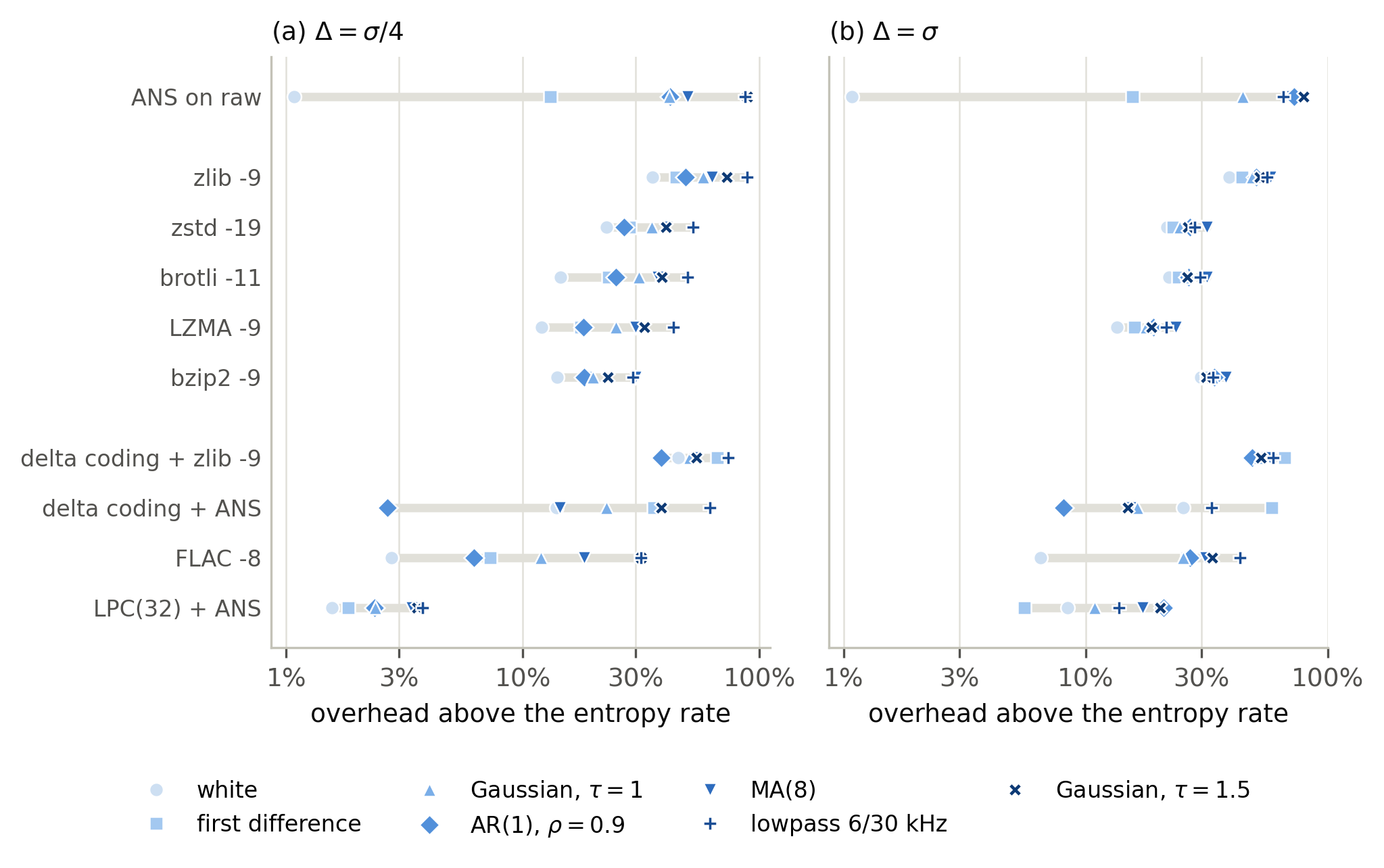}
\caption{Overhead above the entropy rate for every coder and every family, at (a) $\Delta = \sigma/4$ and (b) $\Delta = \sigma$, on a logarithmic scale. Each marker is one family; the gray bar spans the seven. Shading darkens as the family becomes more compressible at a fixed step, that is, as its entropy rate falls.}
\label{fig:10}
\end{figure}

None of the general-purpose compressors comes near the limit; at \(\Delta = \sigma/4\) they run from 12 to 89 percent above the rate, with LZMA and bzip2 the strongest of the five and zlib the worst. Their overhead is largest on the strongly smoothed families, precisely the ones with the most structure to exploit, and at \(\Delta = \sigma/4\) zlib codes six of the seven families to more bits than their memoryless entropy, failing to realize even the gains available to a symbol-by-symbol code. These coders search for repeated byte strings, but the redundancy of a quantized noise process lies in the correlation between neighboring samples rather than in exact repetition, so little of it is within their reach.

Delta coding, the simplest transform that uses that correlation, closes part of the gap but is unreliable. Followed by ANS it ranges from 2.7 percent above the rate on AR(1), for which the first difference is close to the right predictor, to 61.6 percent on the lowpass process, for which it is not. For white noise and the first difference, differencing is the wrong operation outright, since it inflates the innovation variance rather than reducing it, and both delta pipelines cost more than their undifferenced counterparts.

FLAC comes closest of the standard alternatives, 2.8 to 31.5 percent above the rate at \(\Delta = \sigma/4\) and better than every general-purpose compressor on five of the seven families, though bzip2 edges past it on the two most strongly smoothed ones. It does not reach the limit either, and its overhead grows to 6.5 to 43.4 percent at \(\Delta = \sigma\). Its distance from the rate is not a matter of predictor order. At level 8 the reference encoder fits predictors of order up to 12, the streamable-subset limit at these sample rates, and lifting that limit to 32, the maximum the format allows, changed nothing on any of these families. Giving FLAC's coder an already whitened stream locates the rest: on the residual of our order-32 predictor for the lowpass process it spends 2.728 bits per sample, against 2.271 for the same residual coded by ANS and 2.188 for the rate itself. Most of what remains is therefore in the Rice coding of the residuals, which is matched to a two-sided geometric distribution rather than to the near-Gaussian residual of a smoothed process.

Our predictive pipeline comes within 2 to 4 percent of the entropy rate at \(\Delta = \sigma/4\) for every family, from 1.6 percent for white noise to 3.8 percent for the lowpass process, and its performance is similar at the finer steps. The gap widens as the quantization coarsens, reaching roughly 6 to 21 percent at \(\Delta = \sigma\), presumably because prediction from rounded values becomes less effective and the residual retains structure that the memoryless entropy coder cannot exploit. For white noise, where there is nothing to predict, ANS alone on the raw stream is within about one percent of the rate at every step.

The predictor order matters more than might be expected. With the seemingly natural choice of order \(L - 1\), matched to the length of the generating kernel, the MA(8) overhead at \(\Delta = \sigma/4\) is 0.32 bits per sample rather than 0.09, and the first difference behaves similarly. These are the families whose spectra have exact zeros, so their AR representations decay slowly and useful prediction requires an order well beyond the kernel length; at order 32 the excess is largely gone. The smooth families are insensitive to this choice.

The compressibility of these signals lies almost entirely in the correlation structure, so a coder must model that structure to approach the limit. A model-free coder pays the order-0 entropy, nearly twice the rate for the strongly smoothed families; the general-purpose compressors do better but remain far above; FLAC, which models the correlation, comes much closer; and a predictor matched to the process, followed by an entropy coder, comes within a few percent.

\section{Discussion}\label{discussion}

We have presented an analytical approximation to the entropy rate of a uniformly quantized stationary Gaussian process, together with a particle-filter method for estimating the exact rate. The approximation replaces the spectrum \(S\) in the classical high-resolution formula with \(S + \Delta^2/12\), the exact spectral density of the dithered process, and thereby remains finite and accurate for the bandlimited and strongly filtered processes on which the classical formula fails. The estimator follows the route established for the information rates of channels with memory \cite{ArnoldEtAl2006,DauwelsLoeliger2008}, which reduces the rate to the log-likelihood of a filter evaluated on a single long typical sequence; its components (minimum-phase spectral factorization, the hidden Markov structure it induces, and the fully adapted particle filter that the model admits in closed form) are standard, and the contribution is the manner in which they are combined into a practical method that works at all quantization scales.

The method has several limitations, each a candidate for future work. The lattice error in the Riemann-sum step is expressed exactly as a Poisson sum but not quantified; bounding it would turn the approximation into a controlled estimate in the fine-to-moderate regime. The slack in the Gaussian entropy upper bound, namely the per-sample non-Gaussianity of \(X + U\), is likewise not quantified. Note that the second error pushes the formula upward, whereas the sign of the lattice error is not established. In the extreme coarse regime the constant floor is an artifact of the additive-noise model, and the formula should not be used there. The estimator, for its part, needs \(h_0 > 0\): when \(S\) vanishes on a band the moving-average representation of Section 4.2 does not exist and the filter cannot run, so the strictly bandlimited case is served only by the approximation. Its leading behavior there is correct \cite{GeigerKoch2019}, but we have not tested its accuracy at finite \(\Delta\), since the spectra of Section 5 are positive almost everywhere. Finally, the development assumes an exactly Gaussian, stationary process. Measured signals are neither, so the rate computed here is best read as the rate of a Gaussian model fitted to the signal's spectrum, a reference against which a real recording can be compared, rather than the exact rate of that recording. Extending the methods to non-Gaussian sources would narrow this gap. Both methods also take \(S\) as given; on measured data \(S\) is itself estimated, and that error propagates into the rate, so the reference is only as good as the spectral estimate behind it.

Both methods appear to lie above the rate. The estimator converges from above, as discussed in Section 4.4, and the approximation presumably overestimates as well: its Gaussian error acts upward by the argument of Section 3.4, and the measured errors of Section 5 are consistent with an overestimate wherever the filter is converged. The true rate is thus constrained on one side only, and a complementary lower bound would be needed to certify it. However, that the two estimates, which share no machinery, agree to within a few millibits over most of the range of Section 5 is reassuring.

The two methods are complementary. When \(\Delta \lesssim \sigma\) the approximation is accurate to a few millibits and runs in milliseconds, so it is the natural choice. The estimator is needed in the remaining cases: when the quantization is coarse, when the exact rate is wanted rather than a close approximation, and for checking the approximation on a new process, as we have done here.

The estimator also does more than measure the rate. At each step the filter reports the predictive distribution \(P(y_t \mid y_1^{t-1})\), and an arithmetic coder driven by it would compress at close to \(\bar H\), reaching the limit that the practical coders of Section 5.3 fall short of. This comes at a much higher cost than the linear predictor of that section, so its value is as a bound on what is achievable rather than as a codec for routine use.

Together the two methods make the entropy rate computable at any quantization scale, including the strongly filtered regime the classical formula cannot reach. The rate then serves as a reference: for a measured noise process, how many bits per sample it carries, and how close a given lossless coder comes to that limit.

\vspace{6pt}
\noindent\textbf{Author Contributions:} J.M. conceived the study, developed the methods and software, performed the experiments, and wrote the manuscript. The author has read and agreed to the published version of the manuscript.

\medskip
\noindent\textbf{Funding:} This research received no external funding.

\medskip
\noindent\textbf{Data Availability Statement:} The code and the numerical results supporting the findings of this study, including the scripts that regenerate every figure and table from those results, are openly available on Zenodo at \url{https://doi.org/10.5281/zenodo.23598896} (version 1.0.0) and are maintained at \url{https://github.com/magland/entropy-gaussian-process-paper}.

\medskip
\noindent\textbf{Acknowledgments:} The Flatiron Institute is a division of the Simons Foundation. During the preparation of this manuscript and study, the author used Claude Opus (Anthropic; several model versions) for the purposes of developing the software, designing and running the numerical experiments, producing the figures and tables, and drafting and revising the text. The author has reviewed and edited the output and takes full responsibility for the content of this publication.

\medskip
\noindent\textbf{Conflicts of Interest:} The author declares no conflicts of interest.

\appendix

\counterwithin*{table}{section}
\renewcommand{\thetable}{\thesection\arabic{table}}

\section{Numerically stable filter arithmetic}\label{numerically-stable-filter-arithmetic}

When an observation falls in a far tail relative to a particle's prediction, \(\Phi(r) - \Phi(l)\) underflows in double precision, so the filter works in log space throughout using \(\mathrm{log\_ndtr} = \ln\Phi\):
\[\ln\alpha = \mathrm{log\_ndtr}(b) + \ln\!\big(1 - e^{\,\mathrm{log\_ndtr}(a) - \mathrm{log\_ndtr}(b)}\big),\]
evaluating \(\ln(1-e^x)\) as \(\ln(-\mathrm{expm1}(x))\) for \(x > -\ln 2\) and as \(\mathrm{log1p}(-e^x)\) otherwise. Here \((a,b) = (-r,-l)\) when \(l + r > 0\) and \((l,r)\) otherwise, using the reflection \(\Phi(r) - \Phi(l) = \Phi(-l) - \Phi(-r)\). Reflecting on the sign of the midpoint guarantees \(a + b \le 0\), which keeps \(\mathrm{log\_ndtr}\) out of the saturated region where two nearly equal values near \(0\) cancel. The likelihood increment (step 2 of the algorithm) is accumulated with a log-sum-exp.

The truncated-normal draw (step 4) uses the same reflection and the inverse of \(\mathrm{log\_ndtr}\) rather than a generic truncated-normal routine:
\[w = \mathrm{ndtri\_exp}\big(\mathrm{logaddexp}(\ln\Phi(a),\; \ln U + \ln\alpha)\big), \qquad U \sim \mathrm{Unif}(0,1),\]
then undoes the reflection and clips to \([l,r)\) as a guard. Since \(\ln\Phi(a)\) and \(\ln\alpha\) are already available from the weight computation, the entire step costs one \(\mathrm{log\_ndtr}\) pair and one \(\mathrm{ndtri\_exp}\) per particle; these transcendental evaluations dominate the runtime.

\section{Filter collapse and its diagnostic}\label{filter-collapse-and-its-diagnostic}

Under fine quantization (\(h_0 \ll \sigma\), \(\Delta \lesssim h_0\)) the state is nearly determined by the observations. If the particle cloud drifts off by more than a cell, each particle is forced to draw an extreme innovation, which the near-unit-circle inverse filter amplifies into a worse error at the next step. Past some point every particle is wrong simultaneously, and the resulting estimate is invalid rather than merely noisy.

The effective sample size does not detect this failure, since the surviving particles agree with one another whether or not they agree with the data; the ESS therefore remains healthy through a collapse. A diagnostic that does work is based on the per-step likelihood increment itself. A particle whose prediction is \(d\) standard deviations from the observed cell contributes about \(\log(\Delta/h_0) - d^2/2\) nats, so a filter that is tracking the state cannot fall far below \(\log\min(1, \Delta/h_0)\) per step, and a step falling tens of nats below that level indicates a loss of lock rather than an unlucky symbol. Note that the scale term matters: under fine quantization every step is legitimately worth about \(\log(\Delta/h_0)\) nats, so a fixed threshold would produce false alarms. Replicates that lose lock are discarded rather than averaged, and a run with no surviving replicates reports a collapse instead of an estimate. Since the spectral approximation of Section 3 shares no machinery with the filter, it also serves as an independent check on every reported result.

\section{Detailed comparison tables}\label{detailed-comparison-tables}

The tables below record every Monte Carlo point of the comparison of Section 5.2: the particle-filter estimate (mean $\pm$ standard error over the replicates), the analytical approximation, and their difference in millibits per sample, together with the particle count $N$, sequence length $n$, and replicate count $r$ used at each point. An entry marked \emph{discarded} lost lock in the indicated number of replicates and is provisional in the sense of Section 5.1; an entry marked \emph{collapsed} had no surviving replicate. The tables are generated directly from the results files in the accompanying repository.

{\small
\begin{longtable}[]{@{}rrrrrrrr@{}}
\toprule\noalign{}
width $w$ & $\Delta/\sigma$ & $N$ & $n$ & $r$ & estimate (bits/sample) & approximation & error (millibits) \\
\midrule\noalign{}
\endhead
\bottomrule\noalign{}
\caption{\textbf{moving average}. The particle-filter estimate against the analytical approximation (error is approximation minus estimate), with the particle count $N$, sequence length $n$, and replicate count $r$ used at each point.}\label{tab:family1}\\
\endlastfoot
1 & 0.125 & 8000 & 10000 & 16 & 5.0490 $\pm$ 0.0020 & 5.0480 & -1.0 \\
1 & 0.25 & 8000 & 10000 & 16 & 4.0501 $\pm$ 0.0025 & 4.0508 & +0.7 \\
1 & 1 & 8000 & 10000 & 16 & 2.1045 $\pm$ 0.0027 & 2.1048 & +0.3 \\
1 & 2 & 8000 & 10000 & 16 & 1.2422 $\pm$ 0.0022 & 1.2546 & +12.4 \\
2 & 0.125 & 8000 & 10000 & 16 & 4.5795 $\pm$ 0.0025 & 4.5839 & +4.4 \\
2 & 0.25 & 8000 & 10000 & 16 & 3.6202 $\pm$ 0.0027 & 3.6207 & +0.4 \\
2 & 1 & 8000 & 10000 & 16 & 1.8400 $\pm$ 0.0025 & 1.8396 & -0.4 \\
2 & 2 & 8000 & 10000 & 16 & 1.1091 $\pm$ 0.0026 & 1.1208 & +11.7 \\
4 & 0.125 & 8000 & 10000 & 16 & 4.1082 $\pm$ 0.0021 & 4.1099 & +1.7 \\
4 & 0.25 & 8000 & 10000 & 16 & 3.1720 $\pm$ 0.0031 & 3.1725 & +0.4 \\
4 & 1 & 8000 & 10000 & 16 & 1.5355 $\pm$ 0.0032 & 1.5327 & -2.7 \\
4 & 2 & 8000 & 10000 & 16 & 0.9299 $\pm$ 0.0028 & 0.9462 & +16.4 \\
6 & 0.125 & 8000 & 10000 & 16 & 3.8304 $\pm$ 0.0021 & 3.8339 & +3.4 \\
6 & 0.25 & 8000 & 10000 & 16 & 2.9139 $\pm$ 0.0026 & 2.9127 & -1.2 \\
6 & 1 & 8000 & 10000 & 16 & 1.3575 $\pm$ 0.0029 & 1.3585 & +0.9 \\
6 & 2 & 8000 & 10000 & 16 & 0.8138 $\pm$ 0.0035 & 0.8449 & +31.1 \\
8 & 0.125 & 8000 & 10000 & 16 & 3.6390 $\pm$ 0.0024 & 3.6395 & +0.5 \\
8 & 0.25 & 8000 & 10000 & 16 & 2.7333 $\pm$ 0.0025 & 2.7313 & -2.0 \\
8 & 1 & 8000 & 10000 & 16 & 1.2390 $\pm$ 0.0021 & 1.2420 & +3.0 \\
8 & 2 & 8000 & 10000 & 16 & 0.7366 $\pm$ 0.0035 & 0.7777 & +41.1 \\
16 & 0.125 & 30000 & 10000 & 16 & 3.1860 $\pm$ 0.0023 & 3.1789 & -7.1 \\
16 & 0.25 & 30000 & 10000 & 16 & 2.3217 $\pm$ 0.0030 & 2.3090 & -12.7 \\
16 & 1 & 30000 & 10000 & 16 & 0.9858 $\pm$ 0.0021 & 0.9935 & +7.6 \\
16 & 2 & 30000 & 10000 & 16 & 0.5743 $\pm$ 0.0021 & 0.6378 & +63.5 \\
32 & 0.125 & 30000 & 10000 & 16 & 2.7692 $\pm$ 0.0035 & 2.7336 & -35.6 \\
32 & 0.25 & 30000 & 10000 & 16 & 1.9420 $\pm$ 0.0014 & 1.9152 & -26.8 \\
32 & 1 & 30000 & 10000 & 16 & 0.7729 $\pm$ 0.0028 & 0.7951 & +22.2 \\
32 & 2 & 30000 & 10000 & 16 & 0.4380 $\pm$ 0.0036 & 0.5306 & +92.6 \\
64 & 0.125 & 30000 & 10000 & 16 & 2.4681 $\pm$ 0.0100 & 2.3098 & -158.3 \\
64 & 0.25 & 30000 & 10000 & 16 & 1.6527 $\pm$ 0.0052 & 1.5596 & -93.1 \\
64 & 1 & 30000 & 10000 & 16 & 0.5993 $\pm$ 0.0028 & 0.6441 & +44.8 \\
64 & 2 & 30000 & 10000 & 16 & 0.3226 $\pm$ 0.0052 & 0.4516 & +129.0 \\
\end{longtable}
}

{\small
\begin{longtable}[]{@{}rrrrrrrr@{}}
\toprule\noalign{}
correlation $\rho$ & $\Delta/\sigma$ & $N$ & $n$ & $r$ & estimate (bits/sample) & approximation & error (millibits) \\
\midrule\noalign{}
\endhead
\bottomrule\noalign{}
\caption{\textbf{AR(1), 64 taps}. The particle-filter estimate against the analytical approximation (error is approximation minus estimate), with the particle count $N$, sequence length $n$, and replicate count $r$ used at each point.}\label{tab:family2}\\
\endlastfoot
0 & 0.125 & 8000 & 10000 & 16 & 5.0490 $\pm$ 0.0035 & 5.0480 & -1.0 \\
0 & 0.25 & 8000 & 10000 & 16 & 4.0480 $\pm$ 0.0029 & 4.0508 & +2.9 \\
0 & 1 & 8000 & 10000 & 16 & 2.1034 $\pm$ 0.0026 & 2.1048 & +1.4 \\
0 & 2 & 8000 & 10000 & 16 & 1.2401 $\pm$ 0.0032 & 1.2546 & +14.5 \\
0.15 & 0.125 & 8000 & 10000 & 16 & 5.0343 $\pm$ 0.0025 & 5.0317 & -2.6 \\
0.15 & 0.25 & 8000 & 10000 & 16 & 4.0346 $\pm$ 0.0028 & 4.0346 & +0.0 \\
0.15 & 1 & 8000 & 10000 & 16 & 2.0946 $\pm$ 0.0024 & 2.0909 & -3.7 \\
0.15 & 2 & 8000 & 10000 & 16 & 1.2304 $\pm$ 0.0026 & 1.2454 & +15.0 \\
0.3 & 0.125 & 8000 & 10000 & 16 & 4.9824 $\pm$ 0.0017 & 4.9802 & -2.2 \\
0.3 & 0.25 & 8000 & 10000 & 16 & 3.9836 $\pm$ 0.0024 & 3.9835 & -0.0 \\
0.3 & 1 & 8000 & 10000 & 16 & 2.0429 $\pm$ 0.0028 & 2.0472 & +4.4 \\
0.3 & 2 & 8000 & 10000 & 16 & 1.1984 $\pm$ 0.0025 & 1.2169 & +18.5 \\
0.45 & 0.125 & 8000 & 10000 & 16 & 4.8828 $\pm$ 0.0025 & 4.8853 & +2.5 \\
0.45 & 0.25 & 8000 & 10000 & 16 & 3.8915 $\pm$ 0.0023 & 3.8895 & -2.0 \\
0.45 & 1 & 8000 & 10000 & 16 & 1.9687 $\pm$ 0.0017 & 1.9680 & -0.7 \\
0.45 & 2 & 8000 & 10000 & 16 & 1.1496 $\pm$ 0.0016 & 1.1660 & +16.4 \\
0.6 & 0.125 & 8000 & 10000 & 16 & 4.7264 $\pm$ 0.0026 & 4.7272 & +0.8 \\
0.6 & 0.25 & 8000 & 10000 & 16 & 3.7348 $\pm$ 0.0024 & 3.7331 & -1.7 \\
0.6 & 1 & 8000 & 10000 & 16 & 1.8393 $\pm$ 0.0024 & 1.8396 & +0.3 \\
0.6 & 2 & 8000 & 10000 & 16 & 1.0710 $\pm$ 0.0020 & 1.0860 & +15.0 \\
0.75 & 0.125 & 8000 & 10000 & 16 & 4.4572 $\pm$ 0.0031 & 4.4541 & -3.0 \\
0.75 & 0.25 & 8000 & 10000 & 16 & 3.4618 $\pm$ 0.0029 & 3.4640 & +2.2 \\
0.75 & 1 & 8000 & 10000 & 16 & 1.6282 $\pm$ 0.0023 & 1.6298 & +1.6 \\
0.75 & 2 & 8000 & 10000 & 16 & 0.9421 $\pm$ 0.0027 & 0.9614 & +19.3 \\
0.9 & 0.125 & 8000 & 10000 & 16 & 3.8597 $\pm$ 0.0028 & 3.8580 & -1.7 \\
0.9 & 0.25 & 8000 & 10000 & 16 & 2.8854 $\pm$ 0.0018 & 2.8837 & -1.7 \\
0.9 & 1 & 8000 & 10000 & 16 & 1.2294 $\pm$ 0.0016 & 1.2329 & +3.5 \\
0.9 & 2 & 8000 & 10000 & 16 & 0.6920 $\pm$ 0.0042 & 0.7432 & +51.2 \\
0.95 & 0.125 & 8000 & 10000 & 16 & 3.3913 $\pm$ 0.0025 & 3.3869 & -4.4 \\
0.95 & 0.25 & 8000 & 10000 & 16 & 2.4389 $\pm$ 0.0023 & 2.4372 & -1.7 \\
0.95 & 1 & 8000 & 10000 & 16 & 0.9731 $\pm$ 0.0023 & 0.9833 & +10.2 \\
0.95 & 2 & 8000 & 10000 & 16 & 0.5442 $\pm$ 0.0030 & 0.6150 & +70.7 \\
\end{longtable}
}

{\small
\begin{longtable}[]{@{}rrrrrrrr@{}}
\toprule\noalign{}
width $\tau$ & $\Delta/\sigma$ & $N$ & $n$ & $r$ & estimate (bits/sample) & approximation & error (millibits) \\
\midrule\noalign{}
\endhead
\bottomrule\noalign{}
\caption{\textbf{Gaussian kernel}. The particle-filter estimate against the analytical approximation (error is approximation minus estimate), with the particle count $N$, sequence length $n$, and replicate count $r$ used at each point.}\label{tab:family3}\\
\endlastfoot
0.25 & 0.125 & 8000 & 10000 & 16 & 5.0495 $\pm$ 0.0023 & 5.0480 & -1.4 \\
0.25 & 0.25 & 8000 & 10000 & 16 & 4.0530 $\pm$ 0.0021 & 4.0508 & -2.2 \\
0.25 & 1 & 8000 & 10000 & 16 & 2.1055 $\pm$ 0.0029 & 2.1048 & -0.7 \\
0.25 & 2 & 8000 & 10000 & 16 & 1.2372 $\pm$ 0.0029 & 1.2546 & +17.5 \\
0.45 & 0.125 & 8000 & 10000 & 16 & 5.0239 $\pm$ 0.0027 & 5.0274 & +3.5 \\
0.45 & 0.25 & 8000 & 10000 & 16 & 4.0320 $\pm$ 0.0026 & 4.0303 & -1.7 \\
0.45 & 1 & 8000 & 10000 & 16 & 2.0840 $\pm$ 0.0030 & 2.0873 & +3.2 \\
0.45 & 2 & 8000 & 10000 & 16 & 1.2304 $\pm$ 0.0034 & 1.2431 & +12.7 \\
0.65 & 0.125 & 8000 & 10000 & 16 & 4.7684 $\pm$ 0.0023 & 4.7691 & +0.6 \\
0.65 & 0.25 & 8000 & 10000 & 16 & 3.7818 $\pm$ 0.0019 & 3.7756 & -6.3 \\
0.65 & 1 & 8000 & 10000 & 16 & 1.8871 $\pm$ 0.0033 & 1.8872 & +0.1 \\
0.65 & 2 & 8000 & 10000 & 16 & 1.1139 $\pm$ 0.0027 & 1.1286 & +14.7 \\
0.85 & 0.125 & 8000 & 10000 & 16 & 4.2271 $\pm$ 0.0023 & 4.2255 & -1.6 \\
0.85 & 0.25 & 8000 & 10000 & 16 & 3.2672 $\pm$ 0.0029 & 3.2653 & -1.8 \\
0.85 & 1 & 8000 & 10000 & 16 & 1.6155 $\pm$ 0.0015 & 1.6124 & -3.1 \\
0.85 & 2 & 8000 & 10000 & 16 & 0.9853 $\pm$ 0.0030 & 1.0056 & +20.2 \\
1 & 0.125 & 8000 & 10000 & 16 & 3.7481 $\pm$ 0.0016 & 3.7453 & -2.8 \\
1 & 0.25 & 8000 & 10000 & 16 & 2.8880 $\pm$ 0.0026 & 2.8884 & +0.3 \\
1 & 1 & 8000 & 10000 & 16 & 1.4704 $\pm$ 0.0026 & 1.4713 & +1.0 \\
1 & 2 & 8000 & 10000 & 16 & 0.9103 $\pm$ 0.0035 & 0.9434 & +33.2 \\
1.15 & 0.125 & 30000 & 10000 & 16 & 3.3572 $\pm$ 0.0034 & 3.3547 & -2.5 \\
1.15 & 0.25 & 30000 & 10000 & 16 & 2.6065 $\pm$ 0.0029 & 2.6095 & +3.1 \\
1.15 & 1 & 30000 & 10000 & 16 & 1.3613 $\pm$ 0.0018 & 1.3647 & +3.4 \\
1.15 & 2 & 30000 & 10000 & 16 & 0.8552 $\pm$ 0.0021 & 0.8940 & +38.8 \\
1.3 & 0.125 & 30000 & 10000 & 16 & 3.0572 $\pm$ 0.0021 & 3.0569 & -0.3 \\
1.3 & 0.25 & 30000 & 10000 & 16 & 2.3934 $\pm$ 0.0027 & 2.3930 & -0.4 \\
1.3 & 1 & 30000 & 10000 & 16 & 1.2684 $\pm$ 0.0026 & 1.2782 & +9.9 \\
1.3 & 2 & 30000 & 10000 & 16 & 0.7908 $\pm$ 0.0018 & 0.8527 & +61.9 \\
1.5 & 0.125 & 30000 & 10000 & 16 & 2.7434 $\pm$ 0.0035 & 2.7457 & +2.3 \\
1.5 & 0.25 & 30000 & 10000 & 16 & 2.1625 $\pm$ 0.0024 & 2.1646 & +2.1 \\
1.5 & 1 & 30000 & 10000 & 16 & 1.1689 $\pm$ 0.0020 & 1.1847 & +15.8 \\
1.5 & 2 & 30000 & 10000 & 16 & 0.7271 $\pm$ 0.0029 & 0.8068 & +79.7 \\
\end{longtable}
}

{\small
\begin{longtable}[]{@{}rrrrrrrr@{}}
\toprule\noalign{}
cutoff / $f_s$ & $\Delta/\sigma$ & $N$ & $n$ & $r$ & estimate (bits/sample) & approximation & error (millibits) \\
\midrule\noalign{}
\endhead
\bottomrule\noalign{}
\caption{\textbf{lowpass, 33 taps}. The particle-filter estimate against the analytical approximation (error is approximation minus estimate), with the particle count $N$, sequence length $n$, and replicate count $r$ used at each point.}\label{tab:family4}\\
\endlastfoot
0.08 & 0.125 & 130000 & 5000 & 16 & 1.4923 $\pm$ 0.0017 & 1.5311 & +38.8 \\
0.08 & 0.25 & 130000 & 5000 & 16 & 1.2417 $\pm$ 0.0023 & 1.2821 & +40.4 \\
0.08 & 1 & 130000 & 5000 & 16 & 0.7601 $\pm$ 0.0027 & 0.8285 & +68.4 \\
0.08 & 2 & 130000 & 5000 & 16 & 0.4820 $\pm$ 0.0044 & 0.6302 & +148.2 \\
0.12 & 0.125 & 130000 & 5000 & 16 & 1.8912 $\pm$ 0.0026 & 1.9158 & +24.6 \\
0.12 & 0.25 & 130000 & 5000 & 16 & 1.5713 $\pm$ 0.0027 & 1.5940 & +22.7 \\
0.12 & 1 & 130000 & 5000 & 16 & 0.9570 $\pm$ 0.0033 & 0.9926 & +35.5 \\
0.12 & 2 & 130000 & 5000 & 16 & 0.6315 $\pm$ 0.0029 & 0.7261 & +94.6 \\
0.16 & 0.125 & 260000 & 5000 & 16 & 2.2916 $\pm$ 0.0032 & 2.2945 & +2.9 \\
0.16 & 0.25 & 130000 & 5000 & 16 & 1.8840 $\pm$ 0.0023 & 1.8957 & +11.7 \\
0.16 & 1 & 130000 & 5000 & 16 & 1.1221 $\pm$ 0.0024 & 1.1443 & +22.3 \\
0.16 & 2 & 130000 & 5000 & 16 & 0.7425 $\pm$ 0.0033 & 0.8098 & +67.4 \\
0.2 & 0.125 & 130000 & 5000 & 16 & 2.6573 $\pm$ 0.0035 & 2.6658 & +8.5 \\
0.2 & 0.25 & 130000 & 5000 & 16 & 2.1811 $\pm$ 0.0038 & 2.1882 & +7.1 \\
0.2 & 1 & 130000 & 5000 & 16 & 1.2721 $\pm$ 0.0029 & 1.2853 & +13.2 \\
0.2 & 2 & 130000 & 5000 & 16 & 0.8430 $\pm$ 0.0033 & 0.8837 & +40.7 \\
0.25 & 0.125 & 130000 & 5000 & 16 & 3.1159 $\pm$ 0.0027 & 3.1154 & -0.5 \\
0.25 & 0.25 & 130000 & 5000 & 16 & 2.5357 $\pm$ 0.0041 & 2.5413 & +5.6 \\
0.25 & 1 & 130000 & 5000 & 16 & 1.4510 $\pm$ 0.0030 & 1.4492 & -1.8 \\
0.25 & 2 & 130000 & 5000 & 16 & 0.9470 $\pm$ 0.0035 & 0.9651 & +18.1 \\
0.3 & 0.125 & 130000 & 5000 & 16 & 3.5584 $\pm$ 0.0042 & 3.5557 & -2.8 \\
0.3 & 0.25 & 130000 & 5000 & 16 & 2.8842 $\pm$ 0.0034 & 2.8830 & -1.2 \\
0.3 & 1 & 130000 & 5000 & 16 & 1.5987 $\pm$ 0.0029 & 1.6012 & +2.5 \\
0.3 & 2 & 130000 & 5000 & 16 & 1.0227 $\pm$ 0.0030 & 1.0368 & +14.1 \\
0.36 & 0.125 & 30000 & 10000 & 16 & 4.0751 $\pm$ 0.0030 & 4.0697 & -5.4 \\
0.36 & 0.25 & 30000 & 10000 & 16 & 3.2804 $\pm$ 0.0026 & 3.2791 & -1.3 \\
0.36 & 1 & 30000 & 10000 & 16 & 1.7682 $\pm$ 0.0020 & 1.7707 & +2.5 \\
0.36 & 2 & 30000 & 10000 & 16 & 1.1001 $\pm$ 0.0015 & 1.1126 & +12.5 \\
0.45 & 0.125 & 8000 & 10000 & 16 & 4.8154 $\pm$ 0.0029 & 4.8155 & +0.0 \\
0.45 & 0.25 & 8000 & 10000 & 16 & 3.8500 $\pm$ 0.0024 & 3.8489 & -1.1 \\
0.45 & 1 & 8000 & 10000 & 16 & 2.0062 $\pm$ 0.0022 & 2.0030 & -3.2 \\
0.45 & 2 & 8000 & 10000 & 16 & 1.1950 $\pm$ 0.0019 & 1.2100 & +15.0 \\
\end{longtable}
}

{\small
\begin{longtable}[]{@{}lrrrrrrr@{}}
\toprule\noalign{}
process & $\Delta/\sigma$ & $N$ & $n$ & $r$ & estimate (bits/sample) & approximation & error (millibits) \\
\midrule\noalign{}
\endhead
\bottomrule\noalign{}
\caption{\textbf{White noise and first difference}. The particle-filter estimate against the analytical approximation over the quantization step (error is approximation minus estimate).}\label{tab:family5}\\
\endlastfoot
white noise & 0.125 & 8000 & 10000 & 16 & 5.0466 $\pm$ 0.0021 & 5.0480 & +1.4 \\
white noise & 0.25 & 8000 & 10000 & 16 & 4.0456 $\pm$ 0.0030 & 4.0508 & +5.2 \\
white noise & 0.5 & 8000 & 10000 & 16 & 3.0610 $\pm$ 0.0028 & 3.0620 & +1.0 \\
white noise & 1 & 8000 & 10000 & 16 & 2.1094 $\pm$ 0.0025 & 2.1048 & -4.6 \\
white noise & 2 & 8000 & 10000 & 16 & 1.2407 $\pm$ 0.0029 & 1.2546 & +13.9 \\
white noise & 4 & 8000 & 10000 & 16 & 0.3136 $\pm$ 0.0019 & 0.6583 & +344.7 \\
first difference & 0.125 & 8000 & 10000 & 16 & 4.5843 $\pm$ 0.0020 & 4.5839 & -0.4 \\
first difference & 0.25 & 8000 & 10000 & 16 & 3.6205 $\pm$ 0.0027 & 3.6207 & +0.2 \\
first difference & 0.5 & 8000 & 10000 & 16 & 2.6929 $\pm$ 0.0028 & 2.6941 & +1.2 \\
first difference & 1 & 8000 & 10000 & 16 & 1.8423 $\pm$ 0.0030 & 1.8396 & -2.8 \\
first difference & 2 & 8000 & 10000 & 16 & 1.1077 $\pm$ 0.0026 & 1.1208 & +13.1 \\
first difference & 4 & 8000 & 10000 & 16 & 0.2923 $\pm$ 0.0031 & 0.6226 & +330.3 \\
\end{longtable}
}

\section{Implementation}\label{implementation}

Three implementations accompany the paper, all in the repository at \url{https://github.com/magland/entropy-gaussian-process-paper}.

The first is a standalone Python package implementing the full pipeline: process specification by preset, convolution kernel, or autocovariance; the cepstral minimum-phase factorization with its diagnostics; the fully adapted filter with the collapse diagnostic; the analytical approximation; and the compression benchmark. A command-line interface produces the estimates (streaming replicates with a running mean and standard error, with the approximation reported alongside) and the plots.

The second is a WebGPU compute shader. The filter is implemented as a single WGSL kernel, with one workgroup per replicate and particles vectorized within it. The same shader source is executed both from Python (via WebGPU bindings) and natively in the browser. Because WebGPU lacks double precision, the log-space normal CDF and its inverse were rebuilt in single precision; the per-step increments are accumulated in double precision on the host, where the collapse diagnostic is applied unchanged.

The third is an interactive web application: the user adjusts the process (filter) and the quantizer, and the estimated rate, the analytical approximation, and real codec performance are reported in response.

All three implementations, the experiment scripts, and the text of the paper were developed with the assistance of a generative AI model, Claude Opus (Anthropic; several model versions), which was used in writing the software, in designing and running the numerical experiments, in producing the figures and tables, and in drafting and revising the manuscript. The author directed this work and reviewed its output.

\end{document}